\documentclass[twocolumn,english]{revtex4-2}
\usepackage[utf8]{inputenc}
\usepackage[T1]{fontenc}
\usepackage{amsmath}
\usepackage{amsfonts}
\usepackage{amssymb}
\usepackage[version=4]{mhchem}
\usepackage{stmaryrd}
\usepackage{hyperref}
\hypersetup{colorlinks=true, linkcolor=blue, filecolor=magenta, urlcolor=cyan,}
\usepackage{mathrsfs}
\usepackage[dvips]{graphicx}
\usepackage{bm}% bold math
\usepackage{amsthm,amssymb,amsmath,amsfonts,latexsym,enumerate,float}
\usepackage[utf8]{inputenc}
\usepackage{cmap}

\begin{document}
\title{Charge transfer transitions and circular magnetooptics in ferrites}

\author{A.S. Moskvin}
\affiliation{Ural Federal University, 620083 Ekaterinburg, Russia}
\affiliation{M.N. Mikheev lnstitute of Metal Physics of Ural Branch of Russian Academy of Sciences, 620108 Ekaterinburg, Russia}

\begin{abstract}
The concept of charge-transfer (CT) transitions in ferrites is based on the cluster approach and takes into account the relevant interactions as the low-symmetry crystal field, spin-orbital, Zeeman, exchange and exchange-relativistic interactions. For all its simplicity, this concept yield a reliable qualitative and quantitative microscopic explanation of spectral, concentration, temperature, and field dependences of optic and magneto-optic properties ranging from the isotropic absorption as well as the optic anisotropy to the circular magneto-optics. In this review paper, starting with a critical analysis of the fundamental shortcomings of the  "first-principles" DFT-based band theory we present the main ideas and techniques of the cluster theory of the CT transitions to be main contributors to circular magneto-optics of ferrites.
\end{abstract}

%\keywords{Jahn-Teller effect, disproportionation, local composite bosons, spin-triplet superconductivity.}

\maketitle

\section{Introduction}
Over the past 175 years since Michael Faraday's discovery of the relation between light and electromagnetism, magneto-optics has become a broad field of fundamental and applied research. On the one hand, magneto-optics is aimed at the experimental study of the electronic and magnetic structure, magnetic anisotropy, magnetic phase transitions, spin-orbital, exchange and exchange-relativistic effects, and on the other hand, at the search for new materials with high magneto-optical characteristics, improvement and development of new magneto-optical applications.
Various ferrites and, especially, bismuth-substituted iron garnets R$_3$Fe$_5$O$_{12}$ (R = Y, or rare-earth ion) occupy a special place among magneto-optical (MO) materials, being one of the main objects of fundamental research and basic materials for creating various devices of applied magneto-optics from magneto-optical sensors and visualizers, the terahertz isolators, circulators, magneto-optical modulators, optical magnetoelectric sensors, nonreciprocal elements of the integrated optics, to promising applications in high density MO data-storage and low-power consumption spintronic nanodevices.

Rare-earth orthoferrites RFeO$_3$, which have been studied since the 60s of the last century, have attracted and continue to attract the particular attention of researchers for several decades owing to their weak ferromagnetism, remarkable magneto-optical properties, spin-reorientation transitions between antiferromagnetic phases, high velocity of domain walls, and many other properties. Their physical properties remain a focus of considerable research due to promising applications in innovative spintronic devices, furthermore, they contribute to an emerging class of materials, multiferroics with strong magnetoelectric coupling.

%Various ferrites offer an useful platform for exploring the interplay of microwave, spin current, magnetic, and optical degrees of freedom.

The problem of describing the optical and magneto-optical properties of ferrites is one of the most challenging tasks in the theory of strongly correlated 3d compounds. Despite many years of experimental and theoretical research, the nature of their optical and, especially, magneto-optical response remains a subject of debate. This concerns both the identification of electronic transitions responsible for the formation of the main optical and magneto-optical properties and the comprehensive calculation of their contribution to the optical and magneto-optical response functions. The solution of this problem largely depends on the choice of the optimal strategy for taking into account the effects of charge transfer and strong local correlations, which can be formulated as the choice of a compromise between the one-electron band and atomic-molecular description of electronic states.

The nature of the low-energy optical electron-hole excitations in the
insulating transition metal 3d oxides represents one of the most important
challenging issues for these strongly correlated systems.
All these excitations are
especially interesting because they could play a central role in multiband Hubbard
models used to describe both the insulating state
and the unconventional states developed under electron or hole doping. Because of the matrix element effect the optical response does provide only an indirect information about the density of states. Nevertheless it remains one of the most efficient technique to inspect the electronic structure and energy spectrum.

In this review paper, we present a critical analysis of band approaches to describing the optical and magneto-optical response of 3d ferrite-type compounds based on the use of density functional theory (DFT) and argue that the traditional physically transparent atomic-molecular cluster approach (see, e.g.,\,\cite{Kahn,CT-2011,CT-2016} and references therein) based on local symmetry, strong  covalence and charge transfer (CT) effects  with strong local correlations, provides a consistent description and explanation of the optical and magneto-optical response of various ferrites in a wide spectral range.
The review was stimulated by the lack of  detailed and reliable
studies of electron-hole excitations and of a proper understanding
of the relative role of different transitions to optical and  magnetooptical response for ferrites.

The rest of the paper is organized as follows.  In Sec.\, 2 we present a critical overview of the DFT based approaches for description of the optical and magnetooptical properties of strongly correlated 3d compounds and point to the cluster model as a comprehensive physically clear alternative to the DFT approach. In Sec.\,3 we address the charge transfer (CT) states and CT transitions in octahedral [FeO$_6$]$^{9-}$ and tetrahedral [FeO$_4$]$^{5-}$ clusters as basic elements of crystalline and electronic structure for most ferrites. Here we also show that the CT transitions provide an adequate description of the optical spectra for a wide range of ferrites and other 3d oxides. In Sec.\,4 we discuss different interactions for the CT states with a specific focus on so-called exchange-relativistic interactions, in particular, novel "spin-other-orbit" interaction. In Sec.\,5 we analyze the polarisability tensor for the octahedral [FeO$_6$]$^{9-}$ cluster and argue that its contribution to the optical and magnetooptical anisotropy is determined by different interactions in excited states. In Sec.\,6 we overview  different points of a microscopic theory of circular magnetooptics for ferrite-garnets and weak ferromagnets, including Bi-substituted garnets, specific role of the "spin-other-orbit"\, coupling  in weak ferromagnetic ferrites, the temperature dependence of circular magnetooptics, and the role of the 4f-5d transitions in rare-earth ions.  A brief summary is given in Sec.\,7.

\section{Density functional theory  or cluster model?}

\subsection{So-called "ab initio"\, DFT based approaches}

The electronic states in strongly correlated 3$d$ oxides manifest both significant localization and dispersional features. One strategy to deal with this dilemma is to restrict oneself to small many-electron clusters embedded to a whole crystal, then creating model effective lattice Hamiltonians whose spectra may reasonably well represent the energy and dispersion of the important excitations of the full problem. Despite some shortcomings the method did provide a clear physical picture of the complex electronic structure and the energy spectrum, as well as the possibility of a quantitative modeling.

However, last decades the condensed matter community faced an expanding flurry of papers with the so called {\it ab initio} calculations of electronic structure and physical properties for strongly correlated systems such as 3d compounds  based on density functional theory\,\cite{HK,KS}. Only in recent years has a series of papers been published on {\it ab initio} calculations of the electronic structure, optical and magneto-optical spectra of iron garnets (see e.g., Refs.\,\cite{Oikawa,Iori,Nakashima})

However, DFT still remains, in some sense, ill-defined: many of DFT statements were ill-posed or not rigorously proved.
All efforts to account for the correlations beyond LDA (local density approximation) encounter an insoluble problem of double counting (DC) of interaction terms which had just included into Kohn-Sham
single-particle potential.

Most widely used DFT computational schemes start with a "metallic-like" approaches making use of approximate energy functionals, firstly LDA scheme, which are constructed as expansions around the homogeneous electron gas limit and fail quite dramatically in capturing the properties of strongly correlated systems.
The LDA+U and LDA+DMFT (DMFT, dynamical mean-field theory)\,\cite{LDA+U+DMFT} methods are believed to correct the inaccuracies of approximate DFT exchange correlation functionals. The main idea of these computational approaches consists in a selective description of the strongly correlated electronic states, typically, localized $d$ or $f$ orbitals, using the Hubbard model, while all the other states continue to be treated at the level of standard approximate DFT functionals.
At present the LDA+U and LDA+DMFT methods are addressed to be most powerful tools for the investigation of strongly correlated electronic systems, however, these preserve many shortcomings of the DFT-LDA approach.

Usually the values of effective on-site Coulomb parameters $U_{eff} = U - J$, where
$U$ represent the {\it ad hoc} Hubbard on-site Coulomb repulsion parameter and $J$ the intra-atomic Hund’s exchange integral,
are ordinarily determined by seeking a good agreement of the calculated
properties with the experimental results such as band gaps or
oxidation energies. The values of $U_{eff}$ affect strongly the calculated
material properties even in the ground state, so that it is desirable
to find its optimal values, which also depend on the chosen
exchange-correlation functional. Recent studies have attempted to calculate these parameters
directly based on first principles approaches. Nevertheless, the
calculated values differ widely, even for the same ionic state in a
given material, due to a number of factors such as the choice of the
DFT scheme or the underlying basis set. Although it has become a
common practice that a certain $U_{eff}$ value is chosen {\it a priori} during
the setup of a first principles-based calculation, it is also well
known that a certain $U_{eff}$ value may not work definitively for all
calculation methods and DFT schemes. By independently constraining the field on the
Fe atoms at the octahedral and tetrahedral sites in YIG, the authors\,\cite{Nakashima} have obtained
two different values of $U_{eff}$, i.e., 9.8 eV for octa-Fe and 9.1 eV for  tetra-Fe.
These values are considerably different from those used for iron garnets in previous
works, e.g. $U$\,=\,3.5\,eV and $J$\,=\,0.8\,eV\,\cite{Ching} using the orthonormalized
linear combination of atomic orbitals basis set within constrained
LDA approach and $U_{eff}$\,=\,5.7\,eV\,\cite{Xie}, $U$\,=\,4\,eV\,\cite{Iori,Li}.
The Hubbard and Hund’s $U$ and $J$ parameters were chosen as $U_{eff}$\,=\,2.7\,eV for YIG, 4.7\,eV for LuIG, and 5\,eV for Bi-substituted garnet Bi$_x$Lu$_{3-x}$Fe$_5$O$_{12}$\,\cite{Xu}.

Despite many examples of a seemingly good agreement with experimental data (photoemission and inverse-photoemission spectra, magnetic moments,...) claimed by the DFT community, both the questionable starting point and many unsolved and unsoluble problems give rise to serious doubts in quantitative and even qualitative predictions made within the DFT based techniques.

%Some open questions remain, namely the making use of mean-field like approximations, the choice of the localized basis set to define the occupation matrix, the problems with description of the spin or orbitally degenerated states, the symmetry breaking effects, the calculation of the Hubbard U and other correlation parameters, the formulation of the double counting term, the account for the inter-site Coulomb couplings, the failure to describe the excitation spectra, the lack of a clear physical picture of the complex electronic structure, etc.

Strictly speaking, the DFT is designed for description of ground rather than excited states with no good scheme for excitations. Because an excited-state density does not uniquely determine the potential, there is no general analog of the Hohenberg-Kohn functional for excited states. The standard functionals are inaccurate both for on-site crystal field and for charge transfer excitations\,\cite{Burke}. The DFT based approaches cannot provide the correct atomic limit and the term and multiplet structure\,\cite{Ahn,Ravindran}, which is crucial for description of the optical response for 3d compounds. Although there are efforts to obtain correct results for spectroscopic properties depending on spin and orbital density this problem remains as an open one in DFT research.
Clearly, all these difficulties stem from unsolved foundational problems in DFT and are related to fractional charges and to fractional spins. Thus, these basic unsolved issues in the DFT point toward the need for a basic understanding of foundational issues.

%The lack of a firm basis, both on the theoretical side (in particular, the disregard of the N-representability conditions) as well as in the practical side (construction of functionals that do not satisfy the variational principle) places the HKS-DFT, in our opinion, in the domain of semi-empirical theories.

In other words, given these background problems, the DFT based  models should be addressed as semi-empirical approximate ones rather than  {\it ab initio}  theories. M. Levy introduced in 2010  the term DFA to define density functional approximation instead of DFT, which is believed to quite appropriately describe contemporary DFT\,\cite{Kryachko}.

%\subsection{LSDA}

Basic drawback of the spin-polarized approaches to description of electronic structure for spin-magnetic systems, especially in a simple LSDA scheme\,\cite{Oikawa}, is that these start with a  local density functional in the form
$$
{\bf v}({\bf r})=v_0[n({\bf r})]+\Delta v[n({\bf r}),{\bf m}({\bf r})](\bm{\hat\sigma}\cdot \frac{{\bf m}({\bf r})}{|{\bf m}({\bf r})|})\, ,
$$
where $n({\bf r}),{\bf m}({\bf r})$ are the electron and spin magnetic density, respectively, $\bm{\hat\sigma}$ is the Pauli matrix, that is these imply  presence of
a large fictious local {\it one-electron} spin-magnetic field $\propto (v^{\uparrow}-v^{\downarrow})$, where $v^{\uparrow ,\downarrow}$ are the on-site LSDA spin-up and spin-down potentials. Magnitude of the field is considered to be  governed by the intra-atomic Hund exchange, while its orientation does by the effective molecular, or inter-atomic exchange fields. Despite the supposedly spin nature of the field it produces an unphysically giant spin-dependent rearrangement of  the charge density that cannot be reproduced within any conventional technique operating with spin Hamiltonians. Furthermore, a  direct link with the orientation of the field makes the effect of the spin configuration onto the charge distribution to be unphysically large. However, magnetic long-range order has no significant influence on the redistribution of the charge density. In such a case the straightforward application of the LSDA scheme can lead to an unphysical overestimation of the effects or even to qualitatively incorrect results due to an unphysical effect of a breaking of spatial symmetry induced by a spin configuration.
The DFT-LSDA community needed many years to understand such a physically clear point.

%Similar effects cannot be reproduced in frames of any conventional Heisenberg model.

% Indeed, the basic starting points of the current versions of such spin-polarized approaches as the LSDA exclude any possibility to obtain a reliable quantitative estimation of  the spin-dependent electric polarization in multiferroics.

Overall,   the LSDA approach seems to be more or less  justified for a semi-quantitative description of exchange coupling effects for  materials with a classical N\'eel-like collinear magnetic order. However, it can lead to erroneous results for systems and high-order perturbation effects where the symmetry breaking and quantum fluctuations are of a principal importance such as:  i) noncollinear spin configurations, in particular,  in quantum s\,=\,1/2 magnets; ii)  relativistic effects, such as the symmetric spin anisotropy, antisymmetric DM coupling; iii) spin-dependent electric polarization; iv) circular magnetooptical effects.

In general, the  LSDA method to handle a spin degree of freedom is absolutely incompatible with a conventional approach based on  the spin Hamiltonian concept. There are some intractable problems with a match making between the conventional formalism of a spin Hamiltonian and LSDA approach to the exchange and exchange-relativistic effects.
Visibly plausible numerical results for different exchange and exchange-relativistic  parameters reported in many LSDA investigations (see, e.g., Refs.\,\cite{Mazurenko})  evidence only a potential capacity of the LSDA based models for semiquantitative estimations, rather than for reliable  quantitative data.
%It is worth noting that for all of these "advantageous" instances the matter concerns the handling of certain classical N\'eel-like  spin configurations (ferro-, antiferro-, spiral,...) and search for a compatibility with a mapping made with a  conventional quantum spin Hamiltonian. It's quite another matter when one addresses the search of the charge density redistribution induced by a spin configuration as, for instance, in multiferroics. In such a case the straightforward application of the LSDA scheme can lead to an unphysical overestimation of the effects or even to qualitatively incorrect results due to an unphysically strong effect of a breaking of spatial symmetry induced by a spin configuration (see, e.g. Refs.\,\cite{MF} and references therein).

%Strictly speaking, the DFT is designed for description of ground rather than excited states.
%Nevertheless research activity in the condensed matter DFT community is focused on the single-particle excitation properties of the TMOs, in particular, the photoemission spectra and energy gap.

It is rather surprising how little attention has been paid to the DFT based calculations of the  optical properties for the transition metal oxides (TMO). Lets turn to a  recent paper by  Roedl and Bechstedt\,\cite{NiO_optics} on NiO and other TMOs, whose approach is typical for DFT community.
The authors calculated the dielectric function  $\epsilon (\omega )$ for NiO within the DFT-GGA+U+$\Delta$ technique and claim:"The experimental data agree very well with the calculated curves"\, (!?). However, this seeming agreement is a result of a simple fitting when the two model parameters $U$ and $\Delta$ are determined such (U\,=\,3.0, $\Delta$\,=\,2.0\,eV) that the best possible agreement  concerning the positions and intensities of the characteristic peaks in the experimental  spectra is obtained. In addition, the authors arrive at absolutely unphysical conclusion: "The optical absorption of NiO is dominated by intra-atomic $t_{2g} \rightarrow e_g$ transitions"\, (!?).

There are still a lot of people who think  the Hohenberg-Kohn-Sham DFT within the LDA has provided a very successful {\it ab initio} framework to successfully tackle the problem of the electronic structure of materials. However, both the starting point and realizations of the DFT approach have  raised serious questions. The HK "theorem" of the existence of a mythical universal density functional that can resolve everything looks like a way into Neverland, the DFT heaven is probably unattainable. Various DFAs, density functional approximations, local or nonlocal, will never be exact. Users are willing to pay this price for simplicity, efficacy, and speed, combined with useful (but not yet chemical or physical) accuracy\,\cite{Burke,Becke}.

The most popular DFA fail for the most interesting systems, such as strongly correlated oxides, in particular ferrites. The standard DFT approximations over-delocalize the 3d-electrons, leading to highly incorrect descriptions. Some practical schemes, in particular,  DMFT can correct some of these difficulties, but none has yet become a universal tool of known performance for such systems\,\cite{Burke}.

%In a certain sense the cluster based calculations seem to provide a better description of the overall electronic structure of insulating 3$d$ oxides and its optical response than the DFT based band structure calculations, mainly due to a clear physics and a better account for correlation effects (see, e.g., Refs.\,\cite{Eskes,Ghijsen}).

\subsection{Cluster model approach}

At variance with the DFT theory the cluster model approach does generalize and advance  crystal-field and ligand-field theory. The method provides a clear physical picture of the complex electronic structure and the energy spectrum, as well as the possibility of a quantitative modeling.
In a certain sense the cluster calculations might provide a better description of the overall electronic structure of  insulating  3d oxides  than the band structure calculations\,\cite{Ghijsen,Eskes}, mainly  due to a better account for correlation effects, electron-lattice coupling, and relatively weak interactions such as spin-orbital and exchange coupling.
Moreover, the cluster model has virtually no competitors in the description of impurity or dilute systems.
Cluster models do widely use the symmetry for atomic orbitals, point group symmetry, and advanced technique such as Racah algebra and its modifications for point group symmetry\,\cite{Sugano}. From the other hand the cluster model is an actual proving-ground for various calculation technique from simple quantum chemical MO-LCAO (molecular orbital-linear-combination-of-atomic-orbitals) method to a more elaborate LDA\,+\,MLFT (MLFT, multiplet ligand-field theory)\,\cite{Haverkort} approach.
%generalizing the  crystal-field and ligand-field theory has been successfully applied first for explanation of the d-d crystal field transitions. However, last decades the technique was  and p-d and d-d charge transfer transitions. The author ...
%Quantum chemistry approaches, based on solving a simplified Schroedinger equation, within some approximations/assumptions on the structure of the many-body wave function of the system and on the resulting electronic interactions, represent in themselves a very active field of research in the study of correlated systems.
The LDA\,+\,MLFT technique implies a sort of generalization of conventional ligand-field model with the DFT-based calculations.
Haverkort {\it et al.}\,\cite{Haverkort} start by performing a DFT calculation for the proper, infinite crystal using a modern DFT code which employs an accurate density functional and basis set [e.g., linear augmented plane waves (LAPWs)]. From the (self-consistent) DFT crystal potential they then calculate a set of Wannier functions suitable as the single-particle basis for the cluster calculation. The authors compared the theory with experimental spectra (XAS, nonresonant IXS, photoemission spectroscopy) for different 3d oxides  and found overall satisfactory agreement, indicating that their ligand-field parameters are correct to better than 10\%. However, the authors  have been forced to  treat on-site correlation parameter $U_{dd}$ and orbitally averaged (spherical) $\Delta_{pd}$ parameter as adjustable ones.
Despite the involvement of powerful calculation techniques the numerical results of the LDA\,+\,MLFT approach seem to be more like semiquantitative ones.
%In such a situation we should transfer the center of gravity of the cluster approaches more and more to elaboration of physically sound and clear semiquantitative models that are maximally take into account all the symmetry  requirements  on one hand and refer to experiment on the other.
Nevertheless, any comprehensive physically valid description of the electron and optical spectra for strongly correlated systems, as we suggest, should  combine simple physically clear cluster ligand-field analysis with a numerical calculation technique such as LDA+MLFT\,\cite{Haverkort},  and a regular appeal to experimental data.

It is now believed that  the most intensive low-energy electron-hole excitations
 in insulating 3d oxides correspond to the charge transfer (CT) transitions
 while different phonon-assisted crystal field transitions are generally much weaker.
 Namely the CT transitions are considered as a likely source of the optical and
 magneto-optical response of the 3d metal-based oxide compounds in a wide
 spectral range of 1-10 eV, in particular, of the fundamental absorption edge.
 The low-energy dipole-forbidden \emph{d-d} orbital excitations, or crystal field
 transitions, are characterized by the oscillator strengths which are smaller by a factor
$10^2-10^3$ than those for the dipole-allowed \emph{p-d} CT
transitions and usually correspond to contributions to the
dielectric function $\varepsilon^{\prime\prime}$ of the order of 0.001-0.01.

Despite  CT transitions are well established concept in the solid state physics,
 their theoretical treatment remains  rather naive and did hardly progress during last decades.
 Usually it is based on the {\it one-electron} approach with some 2\emph{p}-3\emph{d}
 or, at best, 2\emph{p}$\rightarrow$ 3\emph{d}$\,t_{2\mathrm{g}}$, 2\emph{p}$\rightarrow
$3\emph{d}$\,e_{\mathrm{g}}$ CT transitions in 3\emph{d} oxides. In
terms of the Hubbard model, this is a CT transition from the
nonbonding oxygen band to the upper Hubbard band. But such a
simplified approach to CT states and transitions in many cases
appears to be absolutely insufficient and misleading even for
qualitative explanation of the observed optical and magneto-optical
properties. First, one should generalize the concept of CT
transitions taking into account  the conventional transition between
the lower and upper Hubbard bands which corresponds to an inter-site
\emph{d-d} CT  transition, or intersite transition across the Mott
gap.

Several important problems  are hardly addressed in the  current
analysis of optical spectra, including the relative role of
different initial and final orbital states and respective CT
channels, strong intra-atomic correlations, effects of strong
electron and lattice relaxation for CT states, the transition matrix
elements, or transition probabilities, probable change in crystal
fields and correlation parameters accompanying the charge transfer.

One of the central issues in the analysis of electron-hole
excitations  is whether low-lying states   are comprised of free
charge carriers or excitons. A conventional approach implies that if
the Coulomb interaction is effectively screened and weak, then the
electrons and holes are only weakly bound and move essentially
independently as free charge-carriers. However, if the Coulomb
interaction between electrons and holes is strong, excitons are
believed to form, i.e.\ bound particle-hole pairs with strong
correlation of their mutual motion.

Despite all the shortcomings the cluster models have proven themselves to be reliable  working models for strongly correlated systems such as 3d compounds. These have a long and distinguished history of application in electron, optical and magnetooptical spectroscopy, magnetism, and magnetic resonance. The author with colleagues has successfully demonstrated great potential of the cluster model for description of the $p$-$d$ and $d$-$d$ charge transfer transitions and their contribution to optical and magneto-optical response  in various 3d oxides such as ferrites\,\cite{Fe-1990,Fe-1991,MO-2,MO-3,MO-4,BIG-1991,Fe-1993,BIG-2002,Fe-2009,MnFe-2010}, cuprates\,\cite{Cu-2002,Cu-2003,Cu-2004,Cu-2010,Cu-2019},  manganites\,\cite{Mn-2002,Mn-2010,MnFe-2010}, and nickelates\,\cite{Ni-2012,Ni-2012_2}.

\section{Cluster model: the CT configurations and CT transitions in ferrites}

\subsection{Electronic structure of octahedral [FeO$_6$]$^{9-}$ clusters in ferrites}

 The slightly distorted octahedral [FeO$_6$]$^{9-}$ clusters are
main optical and magneto-optical centers in weak ferromagnetic orthoferrrites RFeO$_3$, hematite $\alpha$ -Fe$_2$O$_3$, borate FeBO$_3$,
cubic antiferromagnetic garnets like Ca$_3$Fe$_2$Ge$_3$O$_{12}$, and,
together
with tetrahedral [FeO$_4$]$^{5-}$   complexes  in other ferrites as well.

 Five Me\,3d and eighteen  oxygen O\,2p atomic
orbitals in octahedral MeO$_6$ complex with  the point
symmetry group $O_h$ form both hybrid Me\,3d-O\,2p
bonding and antibonding $e_g$ and $t_{2g}$ molecular orbitals,
and purely oxygen nonbonding $a_{1g}(\sigma)$, $t_{1g}(\pi)$,
$t_{1u}(\sigma)$, $t_{1u}(\pi)$, $t_{2u}(\pi)$ orbitals (see, e.g., Refs.\cite{Kahn,Sugano,Mn-2002}). Nonbonding
$t_{1u}(\sigma)$ and $t_{1u}(\pi)$ orbitals with the same symmetry
are hybridized due to the oxygen-oxygen O\,2p$\pi$ -
O\,2p$\pi$ transfer. The relative energy position of different
nonbonding oxygen orbitals is of primary importance for the
spectroscopy of the oxygen--3d--metal charge transfer. This is
firstly determined by the bare energy separation $\Delta \epsilon
_{2p\pi \sigma}=\epsilon _{2p\pi }-\epsilon _{2p\sigma}$ between
O\,2p$\pi$ and O\,2p$\sigma$ electrons.

%\begin{center}
\begin{figure}[t]
\centering
\includegraphics[width=8.5cm,angle=0]{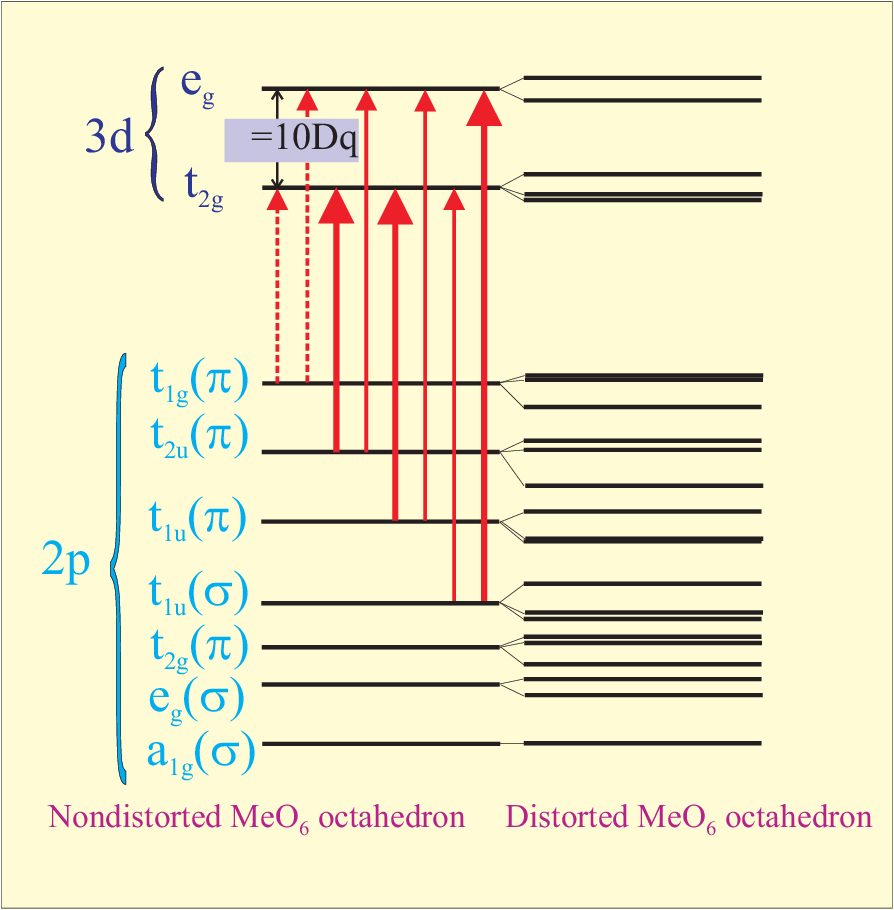}
\caption{The diagram of Me\,3d-O\,2p molecular
orbitals  for the MeO$_{6}$ octahedral center. The O\,2p
- Me\,3d charge transfer transitions are shown by
arrows: strong dipole-allowed  $\sigma -\sigma$  and $\pi -\pi$ by
thick solid arrows;  weak dipole-allowed  $\pi -\sigma$ and $\sigma
-\pi$ by thin solid arrows; weak dipole-forbidden low-energy
transitions by thin dashed arrows, respectively.} \label{fig1}
\end{figure}
%\end{center}

  Since the O\,2p$\sigma$ orbital points towards the two neighboring positive
  3d ions, an electron in this orbital has its energy lowered
   by the Madelung potential as compared with the O\,2p$\pi$ orbitals,
   which are oriented perpendicular
   to the respective 3d--O--3d axes. Thus, Coulomb arguments  favor
    the positive sign of the $\pi -\sigma$ separation
     $\epsilon _{p\pi}-\epsilon _{p\sigma}$ which
  numerical value   can be easily
   estimated in frames of the well-known point charge model, and appears to be of the order of
   $1.0$ eV.
   In a first approximation, all the $\gamma (\pi )$ states
    $t_{1g}(\pi),t_{1u}(\pi),t_{2u}(\pi)$ have the same energy. However,
the O\,2p$\pi$-O\,2p$\pi$ transfer yields the energy
correction
   to bare energies with the largest value
   and positive  sign  for
   the $t_{1g}(\pi)$ state. The energy of the $t_{1u}(\pi)$ state drops due to
   a hybridization with the cation 4\emph{p} $t_{1u}(\pi)$ state. In other words, the $t_{1g}(\pi)$ state is believed
   to be the highest in energy non-bonding oxygen state.
   For illustration, in Figure\,\ref{fig1} we show the energy spectrum of the 3d-2p manifold in the
   octahedral complexes MeO$_6$ with the relative energy position of the levels
   according to the quantum chemical calculations\,\cite{Licht} for the
   [FeO$_{6}$]$^{9-}$ octahedral complex in a lattice environment typical for perovskites
   such as LaFeO$_3$. It should be emphasized one more that the top of the
   oxygen electron band is composed of O\,2p$\pi$ nonbonding orbitals that predetermines
   the role of the oxygen states in many physical properties of 3d perovskites.

%The level scheme of Fig. 1 allows for obvious illustration of the electric polarization effects in MeO$_6$ clusters. First of all one should point to a strong Me3dt$_{2g}$-O2\emph{p}t$_{2u}(\pi)$ and Me3de$_{g}$-O2\emph{p}t$_{1u}(\sigma)$ hybridization effects for Me3dt$_{2g}$ and Me3de$_{g}$ states, respectively, as a source of a big electric polarizability and pseudo-Jahn-Teller (PJT) effect due to a coupling with the odd-parity Q$_{t_{1u}}$ nuclear (core) displacement mode. In the former instance we arrive at the center with a spontaneous dipole moment composed of electronic and nuclear contributions.

The {\it conventional} ground state electronic structure of
octahedral Fe$^{3+}$O$_6$ clusters is associated with the
configuration of the completely filled O\,2p shells and
half-filled Fe\,3d shell. The typical high-spin ground state
configuration and crystalline term for Fe$^{3+}$ in the octahedral
crystal field or for the octahedral [FeO$_{6}$]$^{9-}$ center is
$t_{2g}^{3}e_{g}^2$ and ${}^{6}A_{1g}$, respectively.

The excited CT configuration $\underline{\gamma}_{2p}^1$ 3\emph{d}$^{n+1} \;$
arises from the spin-conserving transition of an electron from the predominantly
anionic molecular orbitals $\gamma_{2p}$  into an empty 3d type MO
($t_{2g}\,$ or $\,e_g$). The transition between the ground
 and the excited configuration can be presented as  the intra-center p-d CT
transition $\,\gamma_{2p}\,\rightarrow\,$ 3d$(t_{2g},e_g)$ .

The p-d CT configuration consists of two partly filledmolecular-orbital subshells, localized predominantly on 3d cation and ligands, respectively. The excited  cation configuration (3d$^6$)  nominally corresponds to the Fe$^{2+}$ ion.
Strictly speaking, the many-electron p-d CT configuration should be written as $t_{2g}^{n_1}e_{g}^{n_2}\underline{\gamma}_{2p}$ with $n_1+n_2=6$, or $((t_{2g}^{n_1}e_{g}^{n_2}){}^{2S^{\prime}+1}\Gamma_{g}^{\prime};\underline{\gamma} _{2p}){}^{2S+1}\Gamma$ ($S = S^{\prime}\pm\frac{1}{2}, \Gamma \in \Gamma_{g}^{\prime} \times \gamma _{2p}$, ${}^{2S+1}\Gamma$ is a crystal term of the CT configuration), if we make use of the spin and orbital quasimomentum addition technique\,\cite{Sugano}.

\subsection{Intra-center electric-dipole p-d CT transitions}

The conventional classification scheme of the intra-center electric-dipole p-d CT transitions in the
octahedral [FeO$_{6}$]$^{9-}$ clusters first of all
includes the electric-dipole allowed transitions from the odd-parity
oxygen
 $\gamma _{u} =t_{1u}(\pi),t_{2u}(\pi),t_{1u}(\sigma)$ orbitals to the even-parity
  iron 3\emph{d}$t_{2g}$ and 3\emph{d}$e_g$  orbitals, respectively.
  These one-electron   transitions
   generate  the many-electron
  ones ${}^{6}A_{1g}\rightarrow {}^{6}T_{1u}$, which  differ by the crystalline
  term of the respective 3\emph{d}$^{6}$ configuration:
\begin{equation}
(t_{2g}^{3}{}^{4}A_{2g};e_{g}^2){}^{6}A_{1g}\rightarrow
((t_{2g}^{4};e_{g}^{2}){}^{5}T_{2g};
\underline{\gamma _{u}}){}^{6}T_{1u},
\end{equation}
\begin{equation}
(t_{2g}^{3}{}^{4}A_{2g};e_{g}^2){}^{6}A_{1g}\rightarrow
((t_{2g}^{3};e_{g}^{3}){}^{5}E_{g};
\underline{\gamma _{u}}){}^{6}T_{1u},
\end{equation}
for $\gamma _{u} \rightarrow$ 3\emph{d}$t_{2g}$ and $\gamma _{u}
\rightarrow$ 3\emph{d}$e_g$ transitions, respectively. We see that
in contrast to the manganese centers Mn$^{3+}$O$_{6}^{9-}$\,\cite{Mn-2002}
 each one-electron $\gamma _{u} \rightarrow$
3\emph{d}$t_{2g}$ transition generates one many-electron CT
transition.

%\subsection{Electric dipole matrix elements}
MeO$_6$
octahedral center can be written with the aid of Wigner-Eckart
theorem\,\cite{Sugano} as follows (see
Ref.\,\cite{Mn-2002} for details)
\begin{equation}
\langle \gamma _{u}\mu|{\hat d}_{q}|\gamma _{g}\mu ^{'}\rangle =
(-1)^{j(\gamma _{u})-\mu}
\left\langle \begin{array}{ccc}
\gamma _{u} & t_{1u} & \gamma _{g} \\
-\mu & q & \mu ^{'}
\end{array}
\right\rangle ^{*} \langle \gamma _{u}\|\hat d\|\gamma _{g}\rangle \,,
\label{d}
\end{equation}
where $\left\langle
\begin{array}{ccc} \cdot & \cdot & \cdot \\ \cdot & \cdot & \cdot
\end{array}
\right\rangle $ is the Wigner coefficient for the cubic point group
O$_h$\,\cite{Sugano}, $j(\Gamma )$ is the so-called quasimomentum number, $\langle
\gamma _{u}\|\hat d\|\gamma _{g}\rangle$ is the one-electron dipole
moment submatrix element. The 3d-2p hybrid structure
of the even-parity molecular orbital
$\gamma_{g}\mu=N_{\gamma_g}(3d\gamma_{g}\mu+\lambda_{\gamma_g}2p\gamma_{g}\mu)$
and a  more simple form of purely oxygen odd-parity molecular
orbital $\gamma_{u}\mu\equiv 2p\gamma_{u}\mu$ both with a symmetry
superposition of the ligand O\,2p orbitals point to a complex
form of  the  submatrix element in  (\ref{d}) to be a sum of $local$
and $nonlocal$ terms composed of the one-site and two-site
(\emph{d-p} and \emph{p-p}) integrals, respectively. In the
framework of a simple "local" approximation that
implies the full neglect of all many-center integrals
$$
\langle t_{2u}(\pi )\|\hat d\|e_{g}\rangle =0;\, \langle t_{2u}(\pi )\|\hat
d\|t_{2g}\rangle = -i\sqrt{\frac{3}{2}}\lambda _{\pi}d \,;
$$
$$
\langle t_{1u}(\sigma )\|\hat d\|t_{2g}\rangle =0;\, \langle t_{1u}(\sigma
)\|\hat d\|e_{g}\rangle = -\frac{2}{\sqrt{3}}\lambda _{\sigma}d \, ;
$$
\begin{equation}
\langle t_{1u}(\pi )\|\hat d\|e_{g}\rangle =0;\, \langle t_{1u}(\pi )\|\hat
d\|t_{2g}\rangle = \sqrt{\frac{3}{2}}\lambda _{\pi}d\, .
\label{d-loc}
\end{equation}
Here, $\lambda _{\sigma}\sim t_{pd\sigma}/\Delta_{pd}$, $\lambda _{\pi}\sim t_{pd\pi}/\Delta_{pd}$
are $effective$ covalency parameters
 for $e_{g},t_{2g}$ electrons, respectively, $d=eR_0$ is an elementary dipole moment
 for the cation-anion bond length $R_0$.
 We see, that the "local" approximation results in an additional selection rule:
 it forbids the $\sigma \rightarrow \pi$, and  $\pi \rightarrow \sigma $
 transitions, $t_{1u}(\sigma )\rightarrow t_{2g}$, and
 $t_{1,2u}(\pi )\rightarrow e_{g}$, respectively, though these are dipole-allowed.
In other words, in frames of this approximation only $\sigma$-type
($t_{1u}(\sigma )\rightarrow e_{g}$) or $\pi$-type ($t_{1,2u}(\pi )\rightarrow
t_{2g}$) CT transitions are allowed.   Hereafter, we make use of the terminology of "strong" and
"weak" transitions for the dipole-allowed CT transitions going on the $\sigma
-\sigma$, $\pi -\pi$, and $\pi -\sigma$, $\sigma -\pi$ channels, respectively. It should be
emphasized that the "local"
approximation, if non-zero, is believed to provide a leading contribution to transition
matrix elements with corrections being of the first order in the cation-anion
overlap integral. Moreover, the nonlocal terms are  neglected in  standard Hubbard-like approaches.
%In Fig.\,\ref{fig2} we do demonstrate the results of numerical calculations of several two-site dipole matrix elements against 3\emph{d} metal - oxygen separation ${\bf R}_{MeO}$.
Given typical cation-anion separations ${\bf R}_{MeO}\approx 4 $ a.u.
we arrive at values less than 0.1 a.u. even for the largest two-site integral, however,
their neglect should be made carefully. Exps.(\ref{d}),(\ref{d-loc}) point to  likely extremely
large dipole matrix elements and oscillator strengths for strong \emph{p-d} CT transitions,
mounting to $d_{ij}\sim$ e\AA\, and $f\sim 0.1$, respectively.

Hence, starting with three  nonbonding purely oxygen orbitals
$t_{1u}(\pi),t_{1u}(\sigma),t_{2u}(\pi)$ as initial states for
one-electron CT, we arrive at six many-electron dipole-allowed CT
transitions ${}^{6}A_{1g}\rightarrow {}^{6}T_{1u}$. There are two
transitions $t_{1u}(\pi),t_{2u}(\pi)\rightarrow t_{2g}$ ($\pi -\pi$
channel), two transitions $t_{1u}(\pi),t_{2u}(\pi)\rightarrow e_{g}$
($\pi -\sigma$ channel), one transition $t_{1u}(\sigma)\rightarrow
t_{2g}$ ($\sigma -\pi$ channel), and one transition
$t_{1u}(\sigma)\rightarrow e_{g}$ ($\sigma -\sigma$ channel).

%In addition to allowed electric dipole transitions, there are a large number of spin- and orbitally-forbidden transitions ${}^6A_{1g}\rightarrow {}^{2S+1}\Gamma_u$ ($S\not=$\,5/2, or $\Gamma_u\not=T_{1u}$) induced by p-d transitions $\gamma_u-3d$ and ${}^6A_{1g}\rightarrow {}^{2S+1}\Gamma_g$  induced by p-d transitions $\gamma_g-3d$ ($\gamma_g$\,=\,$a_{1g}(\sigma)$, $t_{1g}(\pi)$).

It should be noted that the dipole-forbidden
$t_{1g}(\pi)\rightarrow t_{2g}$ transition seemingly  determines the
onset energy of all the p-d CT bands.

\begin{table}
\caption{Parameters (energies, oscillator strength, line width) of the dipole allowed intra-center CT transitions in
octahedral (${}^6A_{1g}\rightarrow {}^6T_{1u}$, No.= 1-6) and tetrahedral (${}^6A_{1g}\rightarrow {}^6T_{2}$, No.= 7-13) clusters in Y$_3$Fe$_5$O$_{12}$\,\cite{Fe-1991,Fe-1993}. E$_{comp}$ and E$_{fit}$ are the computed and fitted CT transition energies, respectively.}
\centering
\begin{tabular}{|c|c|c|c|c|c|}  \hline
No. & Transition
& E$_{comp}$ (eV) & E$_{fit}$ (eV)  & f ($\times\,10^{-3}$)
& $\Gamma$ (eV) \\ [0.1cm] \hline
1 &$t_{2u}\rightarrow t_{2g}$ & 3.1
     & 2.8 &  4 & 0.2 \\
2 &$t_{1u}(\pi)\rightarrow t_{2g}$
& 3.9& 3.6 & 30 & 0.3  \\
3 &$t_{2u}\rightarrow e_g$
& 4.4& 4.3 & 60 & 0.3  \\
4 &$t_{1u}(\sigma)\rightarrow t_{2g}$
& 5.1& 4.8 & 40 & 0.3   \\
5 &$t_{1u}(\pi)\rightarrow e_g$
& 5.3& 5.2 &200 & 0.3  \\
6 &$t_{1u}(\sigma)\rightarrow e_g$
& 6.4& 6.1 &200 & 0.3  \\[0.5cm]
7 &$1t_1\rightarrow 2e$
& 3.4& 3.4 & 30 & 0.4  \\
8 &$6t_2\rightarrow 2e$
& 4.3& 4.6 & 20 & 0.3  \\
9 &$1t_1\rightarrow 7t_2$
& 4.5& 4.7 & 40 & 0.3  \\
10&$5t_2\rightarrow 2e$
& 5.0& 4.9 & 30 & 0.3  \\
11&$6t_2\rightarrow 7t_2$
& 5.4& 5.1 & 20 & 0.3  \\
12&$1e\rightarrow 7t_2$
& 5.6& 5.6 & 10 & 0.3  \\
13&$5t_2\rightarrow 7t_2$
& 6.0& 6.0 & 20 & 0.3  \\
\hline
\end{tabular}
\label{tableCT}
\end{table}

For our analysis to be more quantitative we make two rather obvious
model approximations. First of all, we assume that as usually for
cation-anion octahedra in  3d oxides\,\cite{Kahn,Licht,TMO}
the non-bonding $t_{1g}(\pi)$ oxygen orbital has the highest energy
and forms the first electron removal oxygen state. Furthermore, to
be definite we assume that the energy spectrum of the non-bonding
oxygen states for [Fe$^{3+}$O$_{6}$]$^{9-}$ centers
 coincides with that calculated in Ref.\,\cite{Licht}
for [Fe$^{3+}$O$_{6}$]$^{9-}$ in  orthoferrite LaFeO$_3$,
 in other words, we have (in eV):
 $$
 \Delta (t_{1g}(\pi)-t_{2u}(\pi))\approx 0.8\, ;\,\,
 \Delta (t_{1g}(\pi)-t_{1u}(\pi))\approx 1.8\, ;
 $$
 $$
 \Delta (t_{1g}(\pi)-t_{1u}(\sigma))\approx 3.0\, .
 $$
Secondly, we choose for the Racah parameters $B=0.09$\,eV and $C=0.32$\,eV, the numerical values typical for the Fe$^{3+}$ ion\,\cite{Kahn}.

The energies of the intra-center CT transitions for octahedral FeO$_6$ and tetrahedral FeO$_4$  clusters in Y$_3$Fe$_5$O$_{12}$  were calculated using the spin-polarized X$_{\alpha}$ discrete variational (SP-X$_{\alpha}$ DV) method\,\cite{Fe-1991,Fe-1993}.
These results are presented in Table\,\ref{tableCT} together with the results of fitting the experimental optical data\,\cite{Kahn,Wittekoek}, taking into account only the contribution of the intra-center CT transitions with a Lorentzian line shape.

In addition to several dipole-allowed CT transitions, the CT band will also include various forbidden transitions. First of all, these are dipole-forbidden p-d transitions between states with the same parity of the 2p$t_{1g}$-3d$t_{2g}$ type, as well as satellites of allowed transitions having the same electronic configuration, but different terms of the final
states. For instance in the FeO$_6$-octahedron, these are the $\>^6A_{1g}\,\rightarrow\,^6\Gamma_u\,$ transitions
($\Gamma\,=\,A_1\,,\>A_2\,,\>E\,,\>T_1$) forbidden by the quasimoment
selection rule, and the $\>^6A_{1g}\,\rightarrow\,^4\Gamma_u\,$ spin forbidden
transitions (if $\,\Gamma\, \not =\,T_{1u}\,$, then quasimoment forbidden, too).
The forbiddenness of these transitions is lifted either by the electron-lattice interaction, low-symmetry crystal
field,  spin-orbital interaction, or the exchange interaction with neighbouring clusters. A detailed analysis of the energy spectrum of the CT band requires taking into account the d-d, p-d, and p-p correlation effects.

\subsection{Inter-center \lowercase{\emph{d}-\emph{d}} CT transitions }

Strictly speaking, reliable identification of the intra-center p-d CT transitions is possible only in highly dilute or impurity systems such as YAlO$_3$:Fe or Ca$_3$Fe$_x$Ga$_{2-x}$Ge$_3$O$_{12}$, while in concentrated systems (YFeO$_3$, Ca$_3$Fe$_2$Ge$_3$O$_{12}$, Y$_3$Fe$_5$O$_{12}$, ..) these transitions compete with inter-center d-d CT transitions\,\cite{Cu-2002,Cu-2003,Cu-2004,Cu-2019,Mn-2010,Ni-2012}.

The inter-center \emph{d-d} CT transitions between two MeO$_n$
clusters centered at neighboring sites 1 and 2  define inter-center \emph{d-d} CT excitons in 3d oxides\,\cite{Cu-2002,Cu-2003,Cu-2004,Cu-2019,Mn-2010,Ni-2012}.
These excitons  may be addressed as quanta of the disproportionation
reaction
\begin{equation}
Me_1O_{n}^{v}+Me_2O_{n}^{v}\rightarrow
Me_1O_{n}^{v-1}+Me_2O_{n}^{v+1}\, ,
\label{r1}
\end{equation}
with the creation of electron  \emph{Me}O$_{n}^{v-1}$ and hole
\emph{Me}O$_{n}^{v+1}$ centers. Depending on the initial and final
single particle states all the inter-center \emph{d-d} CT transitions
may be classified to the $e_g-e_g$, $e_g-t_{2g}$, $t_{2g}-e_g$, and
$t_{2g}-t_{2g}$ ones. For  the 3\emph{d} oxides with cations obeying
the Hund rule these can be divided to so-called high-spin (HS)
transitions $S_1S_2S\rightarrow S_{1}\pm \frac{1}{2}S_{2}\mp
\frac{1}{2}S$ and low-spin (LS) transitions $S_1S_2S\rightarrow
S_{1}- \frac{1}{2}S_{2}- \frac{1}{2}S$, respectively.

An inter-center \emph{d-d} CT transition in iron oxides with
Fe$^{3+}$O$_{6}$ octahedra
\begin{equation}
\mbox{[FeO}_{6}]^{9-}+\mbox{[FeO}_{6}]^{9-}\rightarrow
\mbox{[FeO}_{6}]^{10-}+\mbox{[FeO}_{6}]^{8-}
\end{equation}
implies the  creation of electron  [FeO$_{6}$]$^{10-}$ and hole
[FeO$_{6}$]$^{8-}$ centers with electron configurations formally related
to Fe$^{2+}$ and Fe$^{4+}$ ions, respectively. The low-energy inter-center \emph{d-d}
CT transitions from the initial
Fe$^{3+}$O$_{6}(t_{2g}^3e_g^2):{}^6A_{1g}$ states can be directly
assigned to $e_g$$\rightarrow$$e_g$, $e_g$$\rightarrow$$t_{2g}$,
$t_{2g}$$\rightarrow$$e_g$, and $t_{2g}$$\rightarrow$$t_{2g}$ channels
with final configurations and terms
\begin{eqnarray}
e_g\rightarrow e_g:\,
t_{2g}^3e_g^1;{}^5E_g-t_{2g}^3e_g^3;{}^5E_g,\nonumber\\
e_g\rightarrow t_{2g}:\,
t_{2g}^3e_g^1;{}^5E_g-t_{2g}^4e_g^2;{}^5T_{2g},\nonumber\\
t_{2g}\rightarrow
e_g:\,t_{2g}^2e_g^2;{}^5T_{2g}-t_{2g}^3e_g^3;{}^5E_g,\nonumber\\
t_{2g}\rightarrow
t_{2g}:\,t_{2g}^2e_g^2;{}^5T_{2g}-t_{2g}^4e_g^2;{}^5T_{2g}.
\end{eqnarray}
In the framework of high-spin configurations the
$e_g$$\rightarrow$$t_{2g}$ CT transition has the lowest energy
$\Delta =\Delta_{e_g-t_{2g}}$, while  the $e_g$$\rightarrow$$e_g$,
$t_{2g}$$\rightarrow$$t_{2g}$, and $t_{2g}$$\rightarrow$$e_g$
transitions have the energies $\Delta +10Dq(3d^6)$, $\Delta
+10Dq(3d^4)$, and $\Delta +10Dq(3d^6)+10Dq(3d^4)$, respectively. The
transfer energy in the Fe$^{3+}$-based ferrites for the
$e_g$$\rightarrow$$t_{2g}$ CT transition
$$
\Delta^{\mathrm{Fe-Fe}}_{e_gt_{2g}}=A+28\,B-10Dq
$$
can be compared with a similar quantity for the
$e_g$$\rightarrow$$e_g$ CT transition in Mn$^{3+}$-based manganite
LaMnO$_3$
$$ \Delta^{\mathrm{Mn-Mn}}_{e_ge_{g}}=A-8\,B+\Delta_{JT} \, ,
$$
where $\Delta_{JT}$ is the Jahn-Teller splitting of the $e_g$ levels
in manganite. Given $B\approx 0.1$\,eV, $Dq\approx 0.1$\,eV,
$\Delta_{JT} \approx 0.7$\,eV,
$\Delta^{\mathrm{Fe-Fe}}_{e_ge_{g}}\approx 2.0$\,eV (see, e.g., Ref.\,\cite{Kovaleva}) we get $A\approx 2.0$\,eV,
$\Delta^{\mathrm{Fe-Fe}}_{e_gt_{2g}}\approx 4.0$\,eV. In other
words, the onset of the \emph{d-d} CT transitions in Fe$^{3+}$-based
ferrites is strongly ($\sim 2$\,eV) blue-shifted as compared to the
Mn$^{3+}$-based manganite LaMnO$_3$.

Another important difference between ferrites and manganites lies in
the opposite orbital character of initial and final states for the
\emph{d-d} CT transitions. Indeed, the low-energy $d^4d^4\rightarrow
d^3d^5$ CT transition in manganites implies an orbitally degenerate
Jahn-Teller initial state ${}^5E_g{}^5E_g$\,\cite{Bersuker} and an orbitally
nondegenerate final state ${}^4A_{2g}{}^6A_{1g}$ while the
low-energy $d^5d^5\rightarrow d^4d^6$ CT transitions in ferrites
imply an orbitally nondegenerate  initial state
${}^6A_{1g}{}^6A_{1g}$  and an orbitally degenerate Jahn-Teller final
states such as  ${}^5E_g{}^5E_g$ for $e_g\rightarrow e_g$ or
${}^5E_g{}^5T_{2g}$ for $e_g\rightarrow t_{2g}$ CT transitions. An
unconventional final state with an orbital degeneracy on both sites,
or Jahn-Teller excited states  may be responsible for the complex
multi-peak lineshape of the inter-center \emph{d-d} CT band in ferrites.

\subsection{Interplay of the CT transitions in ferrites}

The most complete and detailed analysis of the optical spectra for a wide range of ferrites has been carried out in relatively recent papers\,\cite{Fe-2009,MnFe-2010}. The authors analyze optical
ellipsometry data in the spectral range of 0.6-5.8\,eV for two groups of the iron oxides with more or less distorted FeO$_6$
octahedral and FeO$_4$ tetrahedral clusters. One of the two groups
of materials includes orthoferrites RFeO$_3$, bismuthate BiFeO$_3$,
Y$_{.95}$Bi$_{.05}$FeO$_3$, hematite $\alpha-$Fe$_2$O$_3$,  Fe$_{2-x}$Ga$_x$O$_3$, and borate Fe$_3$BO$_6$ in which iron Fe$^{3+}$ ions
occupy only octahedral centro- or noncentrosymmetric positions and  distortions range from 1 to 20 \%. The second group includes lithium ferrite LiFe$_{5}$O$_{8}$, barium hexaferrite BaFe$_{12}$O$_{19}$, iron garnets R$_{3}$Fe$_{5}$O$_{12}$, and calcium ferrite  Ca$_2$Fe$_2$O$_5$ in which Fe$^{3+}$
ions occupy both octahedral and tetrahedral positions with a rising tetra/ortho ratio. Experimental data were discussed within the cluster model which implies an interplay of intra- (\emph{p-d}) and inter-center (\emph{d-d}) CT transitions.

Some previously reported optical data on ferrites were in most cases
obtained with the use of conventional reflection and absorption
methods. The technique of optical ellipsometry
provides significant advantages over conventional reflection and
transmittance methods in that it is self-normalizing and does not
require reference measurements. The optical complex dielectric
function $\varepsilon=\varepsilon'-i\varepsilon''$ is obtained
directly without a Kramers-Kr\"onig transformation. The dielectric
function $\varepsilon$ was obtained in
the range from 0.6 to 5.8\,eV at room temperature. The comparative
analysis of the spectral behavior of $\varepsilon'$ and
$\varepsilon''$ is believed to provide a more reliable assignement
of spectral features. The spectra were analyzed using the set of the
Lorentz functions

\begin{figure}[t]
\centering
\includegraphics[width=8.5cm,angle=0]{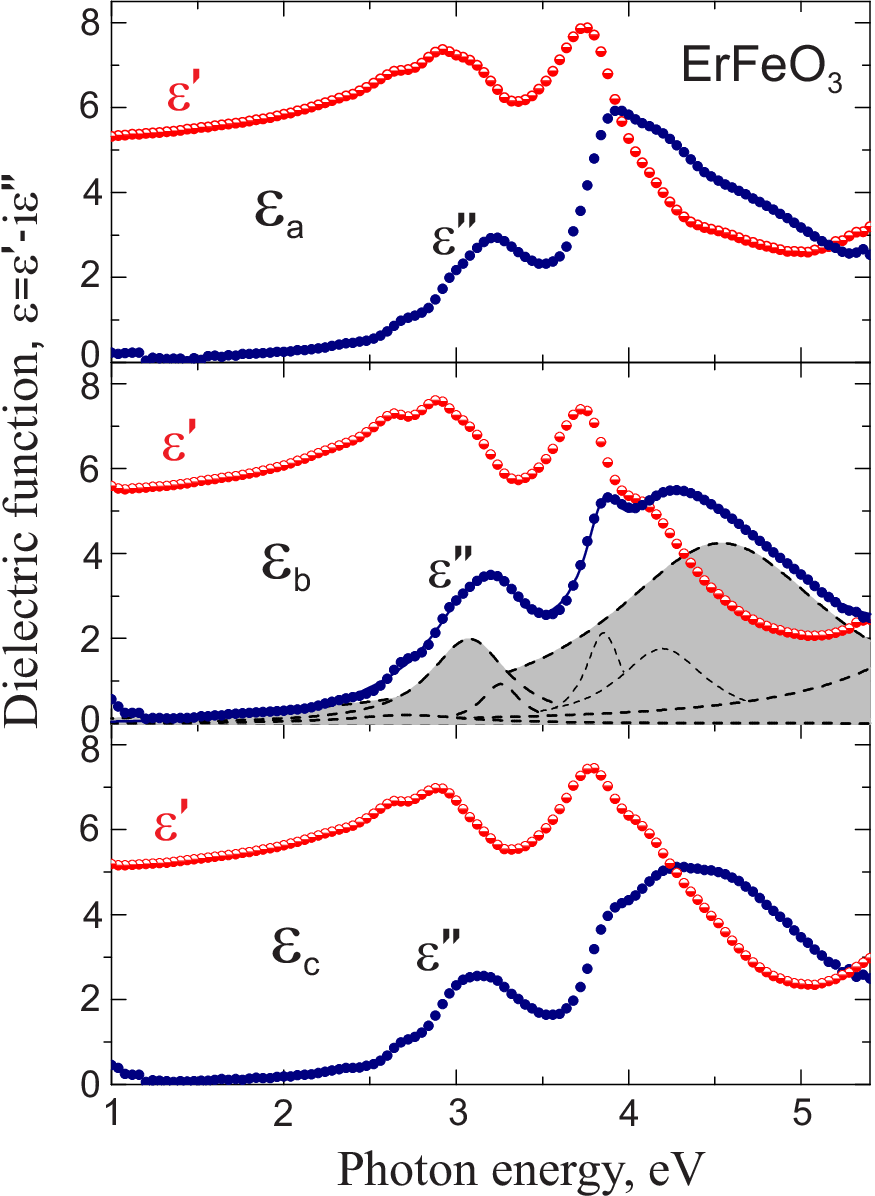}
\caption {(Color online) The dielectric function spectra in ErFeO$_3$ orthoferrite for three  main
polarizations. The Lorentzian fitting is marked by dotted curves and filling.  Insets show indices of absorption and refraction. }
\label{fig2}
\end{figure}

To begin our discussion of the CT transitions in ferrites we refer
to the spectroscopic data for garnets Y$_3$Fe$_x$Ga$_{5-x}$O$_{12}$
(x=5, 3.9, 0.29, 0.09)\,\cite{Scott}. They demonstrate that the
optical response in the spectral range up to 30 000 cm$^{-1}$
($\sim$\,3.7\,eV) is governed by the intra-center transitions for both
octahedral and tetrahedral Fe$^{3+}$ centers. It means that the
onset energy for different \emph{d-d} CT transitions in ferrites is
expected to be $>$\,3.7\,eV in agreement with our model estimates
discussed in Sec.3.3.

To uncover the role played by the octahedral Fe$^{3+}$ centers we
turn to the optical response of the orthoferrites RFeO$_3$.

These compounds contain the only type of centrosymmetric, slightly
($\sim 1\%$) distorted, FeO$_6$ octahedra. Despite the long story of
optical and magneto-optical studies (see, e.g.
Refs.\,\cite{Jung,Kahn}) the microscopic origin of the main
spectral features in orthoferrites remains questionable and the
transition assignments made earlier in Ref.\,\cite{Kahn} need  a
comprehensive revisit. The $\varepsilon', \varepsilon''$  spectra  of ErFeO$_3$ for three main
polarizations  shown in Fig.\,\ref{fig2} are typical for
orthoferrites RFeO$_3$\,\cite{Jung,Kahn,Usachev}. The
low-energy intense band around 3\,eV may be assigned to a strong
dipole allowed intra-center $t_{2u}(\pi)\rightarrow t_{2g}$ CT
transition as  was proposed in Ref.\,\cite{Kahn}. This is a
characteristic feature of the octahedral Fe$^{3+}$ centers in
oxides.
However, such an assignment also implies the existence of a weak
band due to a low-energy dipole-forbidden intra-center
$t_{1g}(\pi)\rightarrow t_{2g}$ CT transition, red-shifted by about
0.8\,eV as expected from estimates\,\cite{Licht}. Indeed, a band
around 2.5\,eV is found in the optical and magneto-optical spectra
of different orthoferrites\,\cite{Kahn}. This band is clearly
visible in hematite $\alpha$-Fe$_2$O$_3$ near 2.4\,eV\,\cite{Fe-2009,MnFe-2010}  where the $t_{1g}(\pi)\rightarrow t_{2g}$
transition becomes allowed due to a breaking of the centro-symmetry
for Fe$^{3+}$ centers.

The nearest high-energy neighborhood of the 3 eV band is expected to
be composed of $t_{1u}(\pi)\rightarrow t_{2g}$ CT transitions with a
comparable intensity and estimated energy about 4\,eV. All the
dipole-allowed intra-center \emph{p-d} CT transitions to the $e_g$
state are blue-shifted by 10$Dq(3d^5)$ as compared to their $\gamma
\rightarrow t_{2g}$ counterparts with the onset energy of the order
of 4\,eV. Interestingly,  for the dipole-allowed $\gamma_u
\rightarrow t_{2g}$ transitions the maximum intensity is expected
for the low-energy $t_{2u}(\pi)\rightarrow t_{2g}$ transition while
for $\gamma_u \rightarrow e_{g}$ transitions the maximum intensity
is expected for the high-energy ($\sim 6-7$\,eV)
$t_{1u}(\sigma)\rightarrow e_{g}$ transition. The analysis of the
experimental spectra for orthoferrites demonstrates the failure of
the intra-center \emph{p-d} CT transitions to explain the broad
intensive band centered near 4.5\,eV together with a narrow
low-energy satellite peaked near 3.9\,eV. Both features are typical
for orthoferrites\,\cite{Jung,Kahn} and may be assigned to a
$e_g\rightarrow t_{2g}$ low-energy inter-center CT transition
${}^6A_{1g}{}^6A_{1g}\rightarrow {}^5E_{g}{}^5T_{2g}$ to an
unconventional final state with an orbital degeneracy on both sites.
These Jahn-Teller excited states  are responsible for the complex
lineshape of the $e_g\rightarrow t_{2g}$ CT band which is composed
of a  narrow exciton-like feature and a broad intense band separated
by $\sim$0.5\,eV, which is believed to be a measure of the
Jahn-Teller splitting in the excited state. Thus we see that all the
spectral features observed in the optical spectra of orthoferrites
for energies below 5\,eV can be directly assigned to the low-energy
intra-center \emph{p-d} and inter-center \emph{d-d} CT transitions.

It is worth noting that the dielectric function in orthoferrites is
nearly isotropic due to very weak ($\sim 1\%$) rhombic distortions
of FeO$_6$ octahedra and nearly equivalent different Fe-O-Fe bonds.
Nevertheless a fine structure of the main CT bands is clearly
revealed in  magneto-optical spectra of orthoferrites, which was
earlier assigned to
 the dipole-forbidden \emph{d-d} crystal field
transitions\,\cite{Jung,Kahn}. In our opinion, their relation to
 the low-symmetry distortions in the \emph{p-d} CT band seems to be more reasonable.

The effect of a strong change in  bulk crystalline symmetry and local trigonal
noncentrosymmetric distortions of FeO$_6$ octahedra is well illustrated by the
optical response of hematite $\alpha$-Fe$_2$O$_3$\,\cite{Fe-2009,MnFe-2010}.
  First of all there is a noticeable rise of intensity and a splitting for dipole-forbidden
$t_{1g}(\pi)\rightarrow t_{2g}$ transition at 2.4\,eV, which is
clearly visible in the spectra of the gallium-substituted sample. Second, one should note a clear splitting
on the order of 0.3-0.4\,eV of the 3\,eV band due to a sizeable
trigonal distortion of the FeO$_6$ octahedra. In both cases the band
splitting effect reflects the singlet-doublet splitting of the
initial orbital triplets, $t_{1g}(\pi)$ and $t_{2u}(\pi)$,
respectively, due to the low-symmetry trigonal crystal field.
Interestingly, the integral intensity of the $t_{2u}(\pi)\rightarrow
t_{2g}$  band at 3\,eV is visibly enhanced in hematite as compared
to similar bands in orthoferrites that may result from the more
covalent Fe-O bonding in hematite.

\section{ Effective Hamiltonian for Fe-clusters in ferrites}

As the principal interactions determining the CT transitions contribution
to the optics and magneto-optics of ferrites, we note the symmetrysymmetry crystal
field $\>(LSCF)\>$, Zeeman interaction $\;V_Z\;$, spin-orbit interaction
$\;V_{SO}\;$, exchange interaction $\;V_{ex}\;$, and the exchange-relativistic
interactions $\;V_{so}^{ex}\>$. The CT configurations have two unfilled
shells -- the $\; 3d^6\,(t_{2g}^4 \, e_g^2 \;$ or $\; t_{2g}^3 \, e_g^3 \,)\;$-
shell and $\; \gamma_{2p} \,$ - shell ($\, \tilde \gamma_{2p}^1 \>$-hole),
which distinguishes them considerably from the ground state configuration having
only one unfilled shell $\; 3d^5 \,(t_{2g}^3 \, e_g^3 \,)\;$ and leads to the
specificity of the manifestation of various interactions, especially anisotropic
ones. Below, we
 consider the aforenamed interactions in the cluster approach.

 \subsection{Low-symmetry crystal field}

Using the the cubic group irreducible tensor operator technique, in particular, the Wigner-Eckart theorem\,\cite{Sugano} we can write the matrix of the  effective Hamiltonian of the low-symmetry crystal field, ${\hat H}_{LSCF},\;$ in general as follows
\begin{widetext}
\begin{equation}
 \langle \kappa SM_s \Gamma M|{\hat H}_{LSCF}| \kappa^{\prime} S^{\prime}M_s^{\prime}\Gamma^{\prime} M^{\prime} \rangle
  =\sum_{\gamma\nu} \sum_{\Gamma\Gamma^{\prime}}  B^{\gamma \star}_{\nu}(\kappa\Gamma\kappa^{\prime}\Gamma^{\prime}) (-1)^{\Gamma - M} \left< \begin{array}{ccc}
  \Gamma & \gamma & \Gamma^{\prime} \\
  -M & \nu & M^{\prime} \\
  \end{array} \right> \,
 \delta_{SS^{\prime}}\delta_{M_s M_s^{\prime}}  \> ,\label{eq:46}
\end{equation}
\end{widetext}
where $\gamma =E,T_2$,  $\;  \langle ::: \rangle \>$  is the $\; 3 \Gamma $ symbol\,\cite{Sugano}, $\kappa ,\kappa^{\prime}$ are certain CT configurations, $B^{\gamma}_{\nu}(\Gamma\Gamma^{\prime})$ are crystal field parameters.

 For a certain $\;T_1 \,(T_2)\;$ term the ${\hat H}_{LSCF}$ can be written as an effective operator
\begin{equation}
V_{LSCF} \> = \> \sum_{ij}\, B_{ij}^{CF}\left[\widetilde{L_iL_j} - \frac{1}{3}L(L\,+\,1)
\,\delta _{ij}\right]\>.\label{eq:20}
\end{equation}
Here, $\; B_{ij}^{CF}\;$ is the symmetric traceless matrix of the $\;LSCF\;$
parameters; $\; \widetilde{L_iL_j}\,=\,(L_iL_j \,+\,L_jL_i)/2\,$, $\; {\bf L}\;$
is the effective orbital moment of the $\;T_1 \,$- ,$\> T_2 \,$- term $\;(L\,=\,1)$.
However, in general, the $\;LSCF\;$ can lead to the mixing of different cubic terms ${}^{2S+1}\Gamma$, ${}^{2S+1}\Gamma^{\prime}$ ($E,T_2\in \Gamma\times\Gamma^{\prime}$)  of identical or different CT configurations with the same spin
multiplicity. All these effects may be of importance, since
$\;H_{LSCF}\;$ reaches the magnitude up to $\; \sim \> $ 0.1 $\> eV\;$ under
the low-symmetry distortions of the [FeO$_6$]$^{9-}$ complex of order
10$^{-2}\,$.

\subsection{Conventional  spin-orbital interaction}

The conventional "intra-center"\, spin-orbital interaction $\;V_{SO}\>=\sum_{i}a(r_i){\bf l}_i\cdot {\bf s}_i \;$
for a certain $\;T_1\,(T_2)\>$ term can be written  as follows
\begin{equation}
V_{so} \> = \> \lambda \,{\bf L} \cdot {\bf S}\> ,   \label{eq:21}
\end{equation}
where $\> \lambda \>$ is the effective spin-orbit coupling constant, tabulated
for the CT states of the [FeO$_6$]$^{9-}$, [FeO$_4$]$^{5-}$
clusters in Refs.\cite{Fe-1991,Fe-1993}. The contributions to $\; \lambda \;$ are due both to the ligand
(oxygen) 2p-subsystem and the iron subsystem, the latter contribution being
dominant. $ \> V_{SO}\;$ leads to the terms splitting and mixing, the latter
being especially significant in case of identical configurations or those
differing from each other in the state of the $\,3d\,$-shell, only. However, in general, the $V_{SO}$ can lead to the mixing of different cubic terms ${}^{2S+1}\Gamma$, ${}^{2S^{\prime}+1}\Gamma^{\prime}$ ($|S-S^{\prime}|\leq 1 \leq S+S^{\prime}; T_1\in \Gamma\times\Gamma^{\prime}$).

\subsection{Zeeman interaction}

The Zeeman interaction $\;V_Z\,=\,\sum_{i} \mu_B ({\bf l}_i \, +
2 {\bf s}_i) \cdot {\bf H} \;$ can be written for a certain $\; T_1 \,(T_2)\>$ term
as an effective operator
\begin{equation}
V_Z \,= \, \mu_B (g_L {\bf L} \> + \> g_S {\bf S}) \cdot {\bf H}\>,\label{eq:22}
\end{equation}
where $\;g_S \;$ and $\; g_L\;$ are respectively the spin $g$-factor
$\;(g_S \approx $ 2)$\;$ and the effective orbital $g$-factor whose
values are listed in Refs.\,\cite{Fe-1991,Fe-1993}. Note, that $\;g_L\;$ can
disagree with the classical orbital value
$\;g_L=1\;$ not only in magnitude, but even in sign. In particular, the CT
state of the $\;t_{2u}^5\,(t_{2g}^4 \, e_g^2 \, \, ^5T_2 \,)\;$ configuration
dominating the magneto-optics of ferrites at the long wavelength tail, has the
value $\;g_L\,=\,-\,\frac{3}{4}\,$. It is worth noting that at variance with the spin-orbital coupling the contributions to $\;g_L\;$
due to the oxygen $\; \tilde \gamma _{2p} \,$-hole and the 3$d$-electrons have comparable values.

\subsection{Exchange interaction}

 The Heisenberg exchange interaction of the [FeO$_6$]$^{9-}$
$m$-cluster in the CT state with the neighbouring $n$-cluster in the ground ${}^6A_{1g}$ state can be written in a simplified form as follows
\begin{equation}
V_{ex}\,=\,-2\, \sum_{m>n}  J_{mn}\left({\bf S}_m\cdot {\bf S}_n\right) \, ,
\label{J}
\end{equation}
where $J_{mn}$ is the exchange integral, although in general it should be replaced by the orbital operator, e.g. for a certain ${}^6T_{1u}$ term for the $m$-cluster
\begin{equation}
\hat{J}_{mn}=J^0_{mn}+\sum\limits_{i=\alpha\beta}J_{mn}^{\alpha\beta}(\widetilde{{\hat L}_{\alpha}{\hat L}_{\beta}}-\frac{2}{3}\delta_{\alpha\beta}) \,.
\label{JJ}
\end{equation}
In general, the cluster spin momentum operators in (\ref{J})  should be replaced by the first rank spin operators, which can change the spin multiplicity. The $V_{ex}$ gives rise to the
orbital and spin splitting and mixing of the $\,CT\,$ configuration terms. The exchange
parameters in $\;V_{ex}\;$ are determined not only by the ordinary cation-anion-cation superexchange
$\;Fe^{3+}$ -- $ O^{2-}$ -- $Fe^{3+}\,$, but also by the considerably stronger direct cation-anion
$\;Fe^{3+}$ -- $ O^{2-}\>$ exchange reaching the magnitude on the order $\,\sim\,0.1\>eV\,$. Strictly speaking, at variance with the antiferromagnetic exchange interaction between the ground states the exchange in the CT state  can lead to both antiferro- and ferromagnetic spin coupling. Interestingly, the matrix of the orbital operator ${\hat J}_{mn}$ in (\ref{JJ}) has a structure similar to $V_{LSCF}$ (\ref{eq:46}) with the main orbitally isotropic $\gamma=A_{1g}$ term included. In other words, nontrivial orbital part of ${\hat V}_{ex}$ can be considered as a spin-dependent contribution to the low-symmetry crystal field.

It should be noted that, in addition to the spin-dependent part, the exchange interaction also contains a spin-independent contribution, which has a similar orbital structure.

\subsection{Exchange-relativistic interactions}

Combined effect of a conventional intra-center spin-orbital coupling and orbitally nondiagonal exchange coupling for an excited orbitally degenerated state of the Fe-cluster within the second-order perturbation theory can give rise to a novel type of exchange-relativistic interaction, modified spin-orbital coupling ${\hat V}_{SO}^{ex}$, which can be written as a sum of isotropic, anisotropic antisymmetric, and anisotropic symmetric intra-center and inter-center terms, respectively\,\cite{Krich,Fe-1990,Fe-1991,Fe-1993}
\begin{widetext}
\begin{equation}
{\hat V}_{SO}^{ex}=\sum_{m,n}\lambda^{(0)}_{mn}({\bf L}_{m}\cdot{\bf S}_{n})+\sum_{m,n}(\pmb{\lambda}_{mn}\cdot[{\bf L}_{m}\times{\bf S}_{n}])+
\sum_{m,n} ({\bf L}_{m}\stackrel{\leftrightarrow}{\pmb{\lambda}}_{mn}{\bf S}_{n}) \, .
\label{SoO}
\end{equation}
\end{widetext}
It is worth noting that $\pmb{\lambda}_{mn}$ has the symmetry of the Dzyaloshinskii vector\,\cite{Dzyaloshinsky,JMMM-2016,CM-2019,JETP-2021}, while the last term  has the symmetry of the two-ion quasidipole spin anisotropy.   Generally speaking, all the three terms can be of a comparable magnitude.

The contribution to the intra-center ($m=n$) bilinear interaction is determined by the spin-independent purely
orbital exchange, while the inter-center ($m\not=n$) term, or "spin-other-orbit"\, coupling ${\hat V}_{SoO}$, is determined by the spin-dependent exchange interaction. However, the spin-dependent exchange leads to
the occurrence of additional nonlinear spin-quadratic
terms, the contribution of which can be taken into
account by the formal replacement of the linear spin
operator ${\bf S}_n$ in (\ref{SoO}) for the nonlinear operator  ${\bf S}_{mn}$

\begin{widetext}
\begin{equation}
\hat S_q(mn)= \hat S_q(n) +\gamma \,\left[\hat V^2 \Bigl( S(m) \Bigr) \, \times \, S^1(n)\right]^1_q  =
 \hat S_q(n)+ \gamma  \sum_{q_1,q_2} \left[\begin{array}{ccc}
2 & 1 & 1 \\
q_1 & q_2 & q
\end{array}\right] \hat V^2_{q_1} \Bigl( S(m) \Bigr) S_{q_2}(n)\>,
\label{Smn}
\end{equation}
\end{widetext}
where [:::] is the Clebsch-Gordan coefficient\,\cite{Sobelman},  $\; V^2_q(S)\;$  is the rank 2 spin irreducible tensor operator. In particular,
\begin{equation}
\hat V_0^2(S)\,=\, 2\,\left[\frac{(2S\,-\,2)!}{(2S\,+\,3)!}\right]^{1/2}
\Bigl( 3 \hat S_z^2 \>-\> S(S \,+\,1) \Bigr)\>.   \label{V2}
\end{equation}
The coefficient $\; \gamma \;$ in (\ref{Smn}) can be calculated for specific terms.
%:  $\;^6 \Gamma _u \,- \;$ and $\;^4\Gamma_u\,-\,$ terms $\;-8\,\sqrt{\frac{6}{7}}\;$ and $\; \sqrt{6 \cdot 7}\;$,
%respectively, for the  terms contributing to the sum over $\;S\Gamma\;$ in
%(\ref{eq:25}).
The isotropic part of $\;V_{SoO}\;$ can be presented, in the
general case, as follows
\begin{widetext}
\begin{equation}
V_{SoO}^{iso} \>= \> \sum_{mn}\, \lambda (mn) \left({\bf L}(m) \cdot {\bf S}(n)\right)
+\sum_{m \not= n}\, \lambda ^{'}(mn) \Bigl( {\bf L}(m) \cdot {\bf S}(m) \Bigr)
\Bigl( {\bf S}(m) \cdot {\bf S}(n) \Bigr)\>.\label{eq:30}
\end{equation}
\end{widetext}

Similarly to the Dzyaloshinskii vector, to estimate the
parameters of the spin-other-orbit coupling, we can
use the simple relation\,\cite{Moriya}
\begin{equation}
\lambda (m) \> \approx \> \lambda (mn) \> \approx \> \frac{\lambda^{\prime} J^{\prime}}
{\Delta E_{S \Gamma}}\> ,\label{eq:31}
\end{equation}
where $\lambda^{\prime}\,$ and $\,J^{\prime}\,$ are the spin-orbital constant for the $T_1$-, $T_2$-states and the nondiagonal exchange parameter, respectively, $\Delta E_{S \Gamma}$ is a certain excitation energy. Parameters like $\lambda (m), \; \lambda (mn)\;$ can be
considerably larger than typical values of the Dzyaloshinskii vector\,\cite{JMMM-2016,CM-2019,JETP-2021}, due both to smaller values
of $\; \Delta E_{S \Gamma}\;$ and to the direct 2$p$--3$d$-exchange which, as
stated above, is stronger than the 3d--2p--3d superexchange determining
$\; {\bf d}(mn)\,$.  Effective orbital magnetic fields acting on the $T_1$ and $T_2$ orbital states, e.g.,
for Fe$^{3+}$ions in ferrites due to $\; V_{SO}^{ex}\;$
can reach the magnitude larger than 10\,T  ($\lambda^{\prime} \geq \>
10^2 \; cm^{-1},\; J^{\prime} \> \geq \> 10^2 cm^{-1}, \; \Delta E_{S\Gamma}
\sim \> 10^4 \> cm^{-1}$).

The approach presented here can be immediately extended to tetrahedral clusters
[FeO$_4$]$^{5-}$.

\section{Anisotropic polarizability of the octahedral  [FeO$_6$]$^{9-}$-cluster}

Almost all ferrites are low anisotropic optical media in a wide spectral range
$\,: \>\Delta \epsilon / \epsilon _0 \leq \,$ 10$^{-2}$,  $\epsilon _0 \;$ and $\; \Delta \epsilon \;$ being respectively
the isotropic and anisotropic parts of the permittivity tensor
$\hat \epsilon$. The latter can be written  as
the sum of the symmetric and antisymmetric parts:
\begin{equation}
\Delta\epsilon \> = \> \Delta\epsilon _{ij}^s \> + \> \Delta\epsilon _{ij}^a\> ,\label{eq:32}
\end{equation}
characterizing the linear birefringence/dichroism  and
the  circular birefringence/dichroism, respectively. The latter can be described by
 axial gyration vector $\> {\bf g}\>$\,\cite{Landau} which is
dual to $\; \Delta\epsilon_{ij}^a \>$ :
\begin{equation}
g_i \> = \> \frac{1}{2} e_{ijk} \Delta\epsilon _{jk}^a \>,    \label{eq:333}
\end{equation}
where $ e_{ijk}$ is the Levi-Civita tensor.

Within a linear approximation the Fe-cluster contribution to anisotropic permitivity tensor can be  expressed in terms of the cluster anisotropic polarizability tensor $\; \hat \alpha \> :$ as follows
\begin{equation}
\Delta\hat \epsilon \> = \> 4 \pi N L \hat \alpha \> ,\label{eq:39}
\end{equation}
where $\;N\;$ is the number of clusters per unit volume; $\; L\,= \,
\frac{n_0^2 \, + \, 2}{9} \;$ is the Lorentz-Lorenz factor. Hence, for the gyration vector we have
\begin{equation}
{\bf g} \> = \> 4 \pi N L {\bm \alpha}\>,\label{eq:40}
\end{equation}
${\bm \alpha}\,$ being the "microgyration vector", related to the antisymmetric
part of the cluster polarizability tensor by an expression analogous to
(\ref{eq:333}).

\subsection{Simple microscopic theory}

The microscopic analysis of the optical anisotropy is usually being carried out on the basis
of the Kramers-Heisenberg formula\,\cite{Condon} for the electronic
polarizability; in case of the microgyration vector it takes on following form :
\begin{equation}
\bm{\alpha}\>=\> \frac{1}{\hbar}\, \sum_{ij}\, \rho_i  \,
 \left[{\bf d}_{ij} \, \times \, {\bf d}_{ji}\right] \cdot F_1(\omega, \, \omega_{ij})\>.
\label{eq:41}
\end{equation}
For the symmetric part of $\; \hat \alpha \>$ , the Kramers-Heisenberg formula
reduces to
\begin{equation}
\alpha_{kl}^{sym} \> = \> \frac{1}{\hbar} \, \sum_{ij} \,\rho_i\,
\langle i |d_k| j \rangle\, \langle j |d_l | i \rangle \cdot F_2(\omega, \,
\omega _{ij})\>.\label{eq:42}
\end{equation}
In these formulae, $\; {\bf d}_{ij} \;$ is the matrix element of the electric
dipole moment $\;{\bf d}\; (d_{k,\, l} \;$ being its Cartesian projections)
between the initial state $\; |i \rangle \;$ and the final state $\; |j\rangle \;$
for the CT transition; $\; \rho_i \>$ is the statistical weight of the
$\; |i \rangle \;$ state. $F_k$ ($k$\,=\,1\,2) is the Lorentz
dispersion factor
\begin{equation}
F_k(\omega, \, \omega_{ij}) \,=\,
\frac{(\omega \, + \, i \Gamma _{ij})[1 \, - \,(-1)^k] \, +
\omega_{ij} \,[1\,+\,(-1)^k]}
{(\omega \,+ \, i \Gamma _{ij})^2 \, - \, \omega _{ij}^2} \;.\label{eq:43}
\end{equation}
Here, $\; \omega _{ij} \;$ denotes the CT transition frequency,
$\; \Gamma _{ij} \;$ is the line width.

Instead of the Cartesian tensor, one can introduce the irreducible polarizability tensor\,\cite{Fe-1991,Fe-1993} :
\begin{equation}
\alpha _q^k \,= \, \frac{1}{\hbar} \sum_{ij} \sum_{q_1 q_2}\, \rho _i
\left[ \begin{array}{ccc}
1 & 1 & k \\
q_1 & q_2 & q
\end{array} \right]
\langle i|d_{q_1}|j \rangle \, \langle j|d_{q_2}|i \rangle
\cdot F_k(\omega, \, \omega_{ij}) \> ,   \label{eq:44}
\end{equation}
where [:::] is the Clebsch-Gordan coefficient\,\cite{Sobelman}, $\; d_q \;$ is the irreducible
tensor component of the dipole moment $\; {\bf d} \;
(d_{\pm 1}\, = \, \mp \frac{1}{\sqrt{2}}(d_x \, \pm \, id_y), \; d_0 \,=\,d_z)$.

An important advantage of the irreducible tensor form is the natural separation of isotropic and anisotropic contributions: $\; \alpha_0^0$
describes the isotropic refraction/absorption; $\; \alpha_q^1\>$  and
$\; \alpha_q^2\>$
describe the circular and linear birefringence/dichroism, respectively.

For octahedral [FeO$_6$]$^{9-}$  (tetrahedral [FeO$_4$]$^{5-}$) clusters with an orbitally nondegenerate ground state ${}^6A_{1g}$ in ferrites, the contribution of the CT transitions ${}^6A_{1g}\rightarrow {}^6T_{1u}$ (${}^6A_{1g}\rightarrow {}^6T_{2}$) to the anisotropic polarizability will be associated only with certain "perturbations" in excited ${}^6T_{1u}$- (${}^6T_{2}$-) CT states.

In the linear approximation, we single out two main contributions $ \alpha _q^k (split)$ and $ \alpha _q^k (mix)$, associated with the orbital splitting of excited ${}^6T$-states and mixing/interaction of different ${}^6T$-states, respectively, under the action of various perturbations, $V_{LSCF}$, $V_Z$, $V_{SO}$, $V_{SO}^{ex}$\,\cite{Fe-1991,Fe-1993}.
\begin{widetext}
\begin{eqnarray}
 \alpha _q^k (split) = \frac{1}{\hbar ^2} \sum_{i=^6A_{1g}}\> \sum_{j=^6T_{1u}}
\> \sum_{\mu \mu ^{'}}\> \sum_{q_1 q_2} \,\rho_i
 \left[ \begin{array}{ccc}
 1 & 1 & k \\
 q_1 & q_2 & q \\
 \end{array}\right]\times        \langle i|d_{q_1} |j \mu \rangle\, \langle j \mu | \hat V |j \mu ^{'} \rangle
\, \langle  j \mu ^{'} |d_{q_2} |i \rangle \cdot \frac{ \partial F_k(\omega,\>\omega_{ij}^0)}
  { \partial \omega _{ij}^{(0)}}          \label{eq:47}
\end{eqnarray}
%\end{widetext}

%\begin{widetext}
\begin{eqnarray}
\alpha _q^k (mix) = \frac{1}{\hbar} \sum_{i=^6A_{1g}} \>
 \sum_{{{j,j^{'}=^6T_{1u}}
\choose {E_j > E_{j^{'}}}}}\> \sum_{q_1 q_2}\, \rho_i
 \left[ \begin{array}{ccc}
 1 & 1 & k \\
 q_1 & q_2 & q \\
 \end{array}\right] \times
\langle i|d_{q_1} |j \rangle \cdot\frac{\langle j|V|j^{'} \rangle}{E_j - E_{j^{'}}}
\cdot\langle j^{'} |d_{q_2} |i \rangle \cdot  F_k(\omega ,\> \omega_{ij} )
\label{eq:48}
\end{eqnarray}
\end{widetext}

\begin{figure}[htbp]
	\centering
	%\sidecaption
		\includegraphics[width=8.5cm,angle=0]{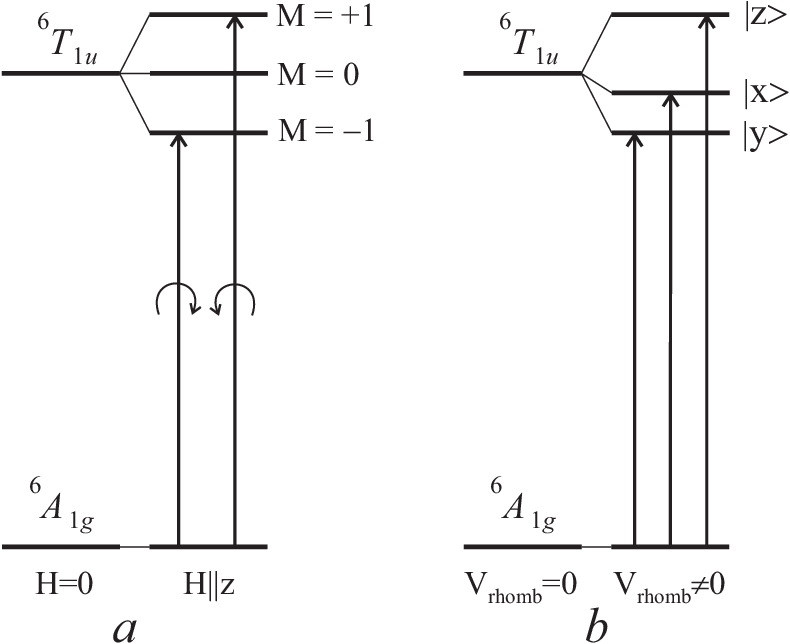}
		\caption{An illustration of the nature of circular and linear birefringence due to a splitting mechanism: ($a$) schematic for the dipole allowed CT transitions $^{6}A_{1g}\rightarrow$ $^{6}T_{1u}$ for the light with right and left circular polarization under external magnetic field and orbital $Zeeman$ splitting;   ($b$) schematic for the CT transitions $^{6}A_{1g}\rightarrow$ $^{6}T_{1u}$ for the light with a linear polarization in a low-symmetry (rhombic) crystal field and $Stark$ splitting for excited $^{6}T_{1u}$ state. Note that we are dealing with finite current (a) and currentless (b) states, respectively. }
	\label{fig3}
\end{figure}

A simple illustration of the nature of circular and linear birefringence due to a splitting mechanism is presented in Figure\,\ref{fig3}.

Note that in ferrites with an orbitally nondegenerate ground ${}^6A_{1g}$ state of Fe-clusters, both linear and circular birefringence will be associated with orbital splitting/mixing in excited states. Obviously, the Fe-cluster contribution to the linear birefringeance/dichroism will be related with low-symmetry crystal field $V_{LSCF}$ in excited ${}^6T_{1u}$ states, while the contribution to circular birefringence/dichroism will be determined by the orbital Zeeman interaction or complex spin-orbital interaction such as $V_{SO}$ and $V_{SO}^{ex}$. Large exchange spin fields up to $10^3$\,T and large spin Zeeman splittings do not make a direct contribution to circular magnetooptics in ferrites.

Due to a competition of the splitting and mixing mechanisms the spectral dependence of the polarizability cannot be considered to be a sum of separate individual ${}^6A{1g}\rightarrow {}^6T$ CT transitions.

\subsection{Symmetry considerations}

Accounting for local point symmetry, crystal and magnetic symmetry in many cases provides important qualitative and even quantitative information about various anisotropic effects, in particular, the role of certain microscopic mechanisms.

\subsubsection{Linear birefringeance in orthoferrites}
Simple symmetry considerations within the framework of the so-called "deformation" model made it possible to explain the dependence of linear birefringence on the type of R-ion in orthoferrites RFeO$_3$\,\cite{M-2021}.

The real FeO$_6$ cluster in orthoferrites can be represented as a homogeneously
deformed ideal octahedron. To find the degree of distortion, we introduce a symmetric strain tensor $\varepsilon_{ij}$ according to the standard rules. In the local system of cubic axes of the octahedron
 \begin{equation}
\varepsilon_{ij}=\frac{1}{4l^2}\sum_{n=1}^6(R_i(n)u_j(n)+R_j(n)u_i(n)) \, ,
\label{u}
 \end{equation}
where ${\bf R}(n)$ is the radius-vector of the Fe-O$_n$ bond, ${\bf u}(n)$ is the O$_n$-ligand displacement vector, or
\begin{equation}
\hat{\varepsilon}=\left(
   \begin{array}{ccc}
    1-\frac{l_1}{l} & \frac{1}{2}(\frac{\pi}{2}-\theta_{12})&\frac{1}{2}(\frac{\pi}{2}-\theta_{13}) \\
	\frac{1}{2}(\frac{\pi}{2}-\theta_{21}) &1-\frac{l_2}{l} & \frac{1}{2}(\frac{\pi}{2}-\theta_{23}) \\
	\frac{1}{2}(\frac{\pi}{2}-\theta_{31})& 	\frac{1}{2}(\frac{\pi}{2}-\theta_{32}) &1-\frac{l_3}{l}	
   \end{array}
  \right) \, ,
  \label{e}
 \end{equation}
 where $l$ is the Fe-O separation in an ideal octahedron, $l_i$ are the Fe-O$_i$ interatomic distances $\frac{1}{3}(l_1+l_2+l_3)=l$, and $\theta_{ij}$ are the bond angles O$_i$-Fe-O$_j$ in a real complex. Local $x,y,z$ axes in octahedron are defined as follows: the $z$-axis is directed along the Fe-O$_I$, the $x$-axis is along Fe-O$_{II}$ with the shortest Fe-O bond length. In general, the deformations of octahedra in orthoferrites are small and do not exceed 0.02.

Diagonal components of the traceless strain tensor (\ref{e}) (tensile/compressive deformations) can be termed as $E$-type deformations since $\varepsilon_{zz}$ and $\frac{1}{\sqrt{3}}(\varepsilon_{xx}-\varepsilon_{yy})$  transform according to the irreducible representation (irrep) $E$ of the cubic group O$_h$, while off-diagonal components (shear deformations) can be termed as $T_2$-type deformations since $\varepsilon_{yz}$,  $\varepsilon_{xz}$,  and  $\varepsilon_{xy}$  transform according to the irrep $T_2$ of the cubic group O$_h$.

In the linear approximation, the symmetric anisotropic polarizability of the octahedron FeO$_6$  can be related to its deformation by the following relation
\begin{equation}
\alpha_{ij}\,=\, \left\{
\begin{array}{cc}
p_E\,    \varepsilon_{ij}\,,&\>i\,=\,j \\
p_{T_2}\,\varepsilon_{ij}\,,&\>i\, \not =\,j\>,
\end{array} \right. \label{eq:65}
\end{equation}
where $\,\varepsilon_{ij}\,$ is the FeO$_6$-octahedron deformation tensor
($Tr\,\hat \varepsilon \,=\,$0); $\>p_{E,\,T_2}\,$ are the photoelastic constants, relating
the polarizability to $E\,,T_{2}$ -deformations, respectively. The relation
(\ref{eq:65}) is
valid in the local coordinate system of the FeO$_6$-octahedron.
In the $\,abc\,$-axes system, it can be rewritten as
\begin{equation}
\alpha_{ij}\>=\>p_E\, \varepsilon_{ij}^E\>+\>\,p_{T_2} \varepsilon_{ij}^{T_2}\>,\label{eq:66}
\end{equation}
where $\,\varepsilon_{ij}^E\,$ and $\,\varepsilon_{ij}^{T_2}\,$ are the components of the
tensor of the $\,E\,$- and $\,T_2\,$-deformations of the octahedron in the $\,abc\,$-
system, respectively.

Proceeding to the permittivity tensor $\,\hat \epsilon\,$ and summing over all
Fe-ions sites, we arrive at nonzero diagonal components of $\,\hat \epsilon\,$:
\begin{equation}
\epsilon_{ii}\>=\>P_E \varepsilon_{ii}^E \>+\>P_{T_2}\varepsilon_{ii}^{T_2}\>,\label{eq:67}
\end{equation}
where $\,P_{E,T_2}\,=\,4\pi N \Bigl(\frac{n_0^2\,+\,2}{3}\Bigr)^2p_{E,T_2}\,$;
$\,\>N\,$ is the number of Fe$^{3+}$ ions per 1\,cm$^3$. Components of
$\,\hat \varepsilon^E,\> \hat \varepsilon^{T_2}\,$ tensors serve as the {\it structure factors}
and may be calculated taking into account the known components of the tensor of
FeO$_6$ octahedron local deformations and the Eulerian angles relating the local axes to the $\,abc\,$ ones.

\begin{figure}[htbp]
	\centering
	%\sidecaption
		\includegraphics[width=8.5cm,angle=0]{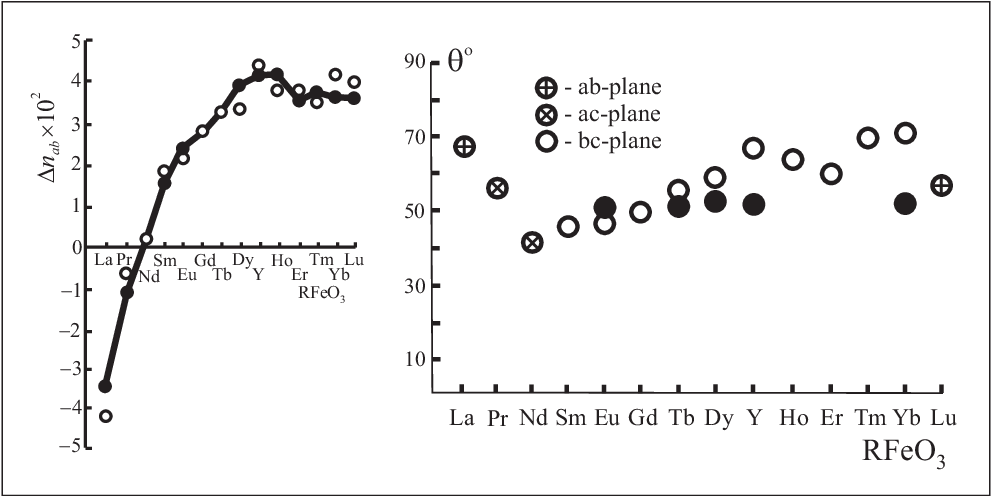}
		\caption{\footnotesize Left panel: Linear birefringeance $\Delta n_{ab}$ for orthoferrites RFeO$_3$ in $ab$-plane, solid circles are predictions of the deformation model, hollow circles are experimental data ($\lambda = 0.633\,\mu m$)\,\cite{Clover}. Right panel: The orientation angles ($\pm\theta$) of optical axes in respective planes of orthoferrites predicted by the deformation model. The solid black circles are scarce experimental data for $bc$-plane ($\lambda = 0.68\,\mu m$)\,\cite{Tabor,Chetkindi}. }
	\label{fig4}
\end{figure}

Thus, we have a two-parameter formula
(\ref{eq:67}) for the birefringence of orthoferrites as a function of rhombic
distortions of their crystal structure.
The photoelastic constants $\,P_E,\> P_{T_2}\,$
can be found from the comparison of experimental data\,\cite{Clover,Tabor} with
the theoretical structure dependence of the $\,ab\,$-plane birefringence :
\begin{equation}
\Delta n_{ab}\,=\,n_{a}-n_{b}\,=\,\frac{1}{2n_0}\left[P_E(\varepsilon_{xx}^E\,-\,\varepsilon_{yy}^E)\,+\,
P_{T_2}(\varepsilon_{xx}^{T_2}\,-\,\varepsilon_{yy}^{T_2})\right]    \label{eq:68}
\end{equation}
treated as a dependence on the type of the orthoferrite. The Figure\,\ref{fig4} shows both
experimental and calculated $\,\Delta n_{ab}\,$  given $\,P_E\,=\,6.2\,n_0\,,
\>P_{T_2}\,=\,4.0\,n_0\,$ (values obtained from the least-squares fitting).
A very nice agreement of the two-parameter formula (\ref{eq:68}) with experiment
testifies to the validity of the deformation model of the birefringence.

Using the found parameter $\,P_{E,T_2}\,$  values, we are able to describe all the
peculiarities of the orthoferrite birefringence. In particular, Figure\,\ref{fig4} shows the theoretical predictions for the orientation angles $\pm\theta$ of optical axes, measured from the $c$-axis for the $ac$- and $bc$-planes and from the $a$-axis for the $ab$-plane, together with  scarce
experimental data on Eu,\,Tb,\,Dy,\,Y,\,Yb orthoferrites\,\cite{Tabor,Chetkindi}. Quite good agreement with the available experimental data is another confirmation of the validity of the deformation model of birefringence of orthoferrites. In general, for all its simplicity, the deformation model reflects quite correctly the main peculiarities of the natural birefringence of orthoferrites.

\subsubsection{Circular birefringeance/dichroism in ferrites}

The gyration vector and the magnetic moment (or the ferromagnetic vector
$\,{\bf m}\,$) have the same transformation properties.

For $\,ferrimagnetic\,$ iron garnets
\begin{equation}
{\bf g}\,=\, \hat A_a{\bf m}_a \, +
\, \hat A_d{\bf m}_d \,+\, \hat C {\bf H} \, ,
\end{equation}
where ${\bf m}_a$ and ${\bf m}_d$ are magnetic moments, or ferromagnetic vectors, of  octahedral and tetrahedral sublattices, respectively.

In weak ferromagnets
like RFeO$_3$ and in a number of other magnetic compounds with
non-equivalent magnetic sublattices, certain components of the ferromagnetic
vector $\;{\bf m}\;$ and the antiferromagnetic vector $\;{\bf l}\;$ in a two-sublattice model transform
identically, what enables one to write $\;{\bf g}\;$ in the linear approximation
through $\,{\bf m},\> {\bf l}\,$, and the external magnetic field $\;{\bf H}\;$
as
\begin{equation}
{\bf g} \> = \> \hat A \,{\bf m} \> + \> \hat B\,{\bf l}\>+\> \hat C\,{\bf H} \, ,
\quad (m^2 \, + \, l^2 \, = \, 1)  \label{eq:35}
\end{equation}
(the ferromagnetic $\>(FM)\,$, antiferromagnetic $\>(AFM)\,$, and field
contributions, respectively)

The form of each of $\,\hat A, \> \hat B, \> \hat C\,$ tensors is determined
by the crystal symmetry. For example, in orthorhombic weak ferromagnetic
orthoferrites RFeO$_3$
\begin{displaymath}
\hat A \>= \>
\left( \begin{array}{ccc}
a_{xx} & 0 & 0 \\
0 & a_{yy} & 0 \\
0 & 0 & a_{zz}
\end{array} \right),    \quad
\hat B \>= \> \hat B^{s} \>+ \> \hat B^{a} \> = \>
\left( \begin{array}{ccc}
0 & 0 & b_{xz} \\
0 & 0 & 0 \\
b_{zx} & 0 & 0
\end{array} \right)\> ,
\end{displaymath}
\[
a_{xx}\not= a_{yy}\not= a_{zz}\, , \quad b_{zx} \not= b_{xz}\>.
\]
In rhombohedral weak ferromagnets $\;(\alpha$-$Fe_2O_3\,,\;FeBO_3\,,\;
FeF_3 \, , \;$ etc.)
\begin{displaymath}
\hat A \> = \>
\left( \begin{array}{ccc}
a_{\perp} & 0 & 0 \\
0 & a_{\perp} & 0 \\
0 & 0 & a_{\parallel}
\end{array} \right),\quad
\hat B \> = \> \hat B^{a} \> = \>
\left( \begin{array}{ccc}
0 & b_{xy} & 0 \\
b_{yx} & 0 & 0 \\
0 & 0 & 0
\end{array} \right)\>,
\end{displaymath}
i.e., $\; b_{yx} = \> -\,b_{xy}\;$, and the $\,\hat B\,$ tensor, in contrast
with orthoferrites, is antisymmetric. The symmetry properties of the
$\; \hat A \;$ and $\; \hat C \;$ tensors are identic.

The special role of the antiferromagnetic contribution to the gyration vector for weak ferromagnets is due to the fact that for them, as a rule, $\; m \ll l \;$, for example, $m/l\approx0.01$ in YFeO$_3$ and $m/l\approx0.001$  in $\alpha$-Fe$_2$O$_3$, respectively\,\cite{CM-2019,JETP-2021,Jacobs}.
However, the components
of the gyration vector $\;{\bf g}\;$ in $\;\alpha$-$Fe_2O_3\;$ and YFeO$_3$ are comparable in
magnitude with  those for the yttrium iron garnet, Y$_3$Fe$_5$O$_{12}$
\cite{Kahn,Zubov1}
although the magnetization of the latter is approximately by two orders larger
than in the hematite and by one order larger than in orthoferrites. It seems
impossible to explain this phenomenon other than in terms of the $\; AFM \;$
contribution.
%It is for the first time experimentally shown in Ref.\cite{Krizub} that the Kerr effect in the hematite is determined almost exclusively by the $\,AFM \,$ contribution.
Hence, it appears that there must be microscopic mechanisms
causing the antisymmetric relations of the gyration vector to spins :
\begin{equation}
{\bf g}\>=\>\sum_{mn}\,\left[{\bf B}(mn)\,\times\,\langle {\bf S}(n)\rangle
\right]\>,
\label{eq:36}
\end{equation}
where the vector $\; {\bf B}(mn) \;$ is determined by the antisymmetric part of
$\> \hat B \,$.

 \section{Charge transfer transitions and magneto-optical effects (MOE) in ferrites}

%The iron garnets  offer an useful platform for exploring the interplay of microwave, spin current, magnetism, and optics degrees of freedom,they have promising applications in high density MO data-storage and low-power consumption spintronic nanodevices.

\subsection{Working microscopic models for circular MOE}

   The main contribution to the microgyration vector for [FeO$_6$]$^{9-}\;$
and [FeO$_4$]$^{5-}$ clusters and the circular MOE for ferrites is determined by the splitting and mixing mechanisms\,\cite{Fe-1991}. To the first order of the perturbation
theory,
only the interactions $\,V_{SO}\,,\;V_Z\,,\;V_{SO}^{ex}\,$ play part, as these are odd in the orbital moment and enable the orbital splitting and
mixing of excited CT states of the $\;^6T_{1u}\;$ type. Note that the spin
part of $\;V_Z\;$ just as the isotropic Heisenberg spin exchange of the
[FeO$_6$]$^{9-}$
cluster with its magnetic surroundings, characterized by the spin exchange
field $\;H_{ex} \;$,
do not contribute in the linear approximation to the circular MOE.
$\;V_{SO}\;$ and the orbital part of $\,V_Z\,$ yield the FM and field
contributions
to the gyration vector; their combined action for the "octahedral" CT
transitions due to the splitting of the excited $\; ^6T_{1u}\;$  states is given by
\begin{equation}
{\bf g}_a^{split}\,=2\,\sum_{j=^6T_{1u}}\,\frac{\pi\,e^2\,L\,N}{\hbar\,m_e\,\omega_{0j}}
\left(\lambda ^j \langle {\bf S} \rangle + \mu_B\, g_L^j\,{\bf H}\right)\,f_j\,
\frac{\partial F_1(\omega,\,\omega_{0j})}{\partial\omega_{0j}}\>.
\label{eq:51}
\end{equation}
where $\langle {\bf S} \rangle$ is the thermodynamic spin average, $f_j$ is the oscillator strength for ${}^6A_{1g}-{}^6T_{1u}$ CT transition, $\lambda ^j$ and $g_L^j$ are effective spin-orbital constant and orbital $g$-factor for a certain ${}^6T_{1u}$ term (see tables\,1 and 2 in Ref.\,\cite{Fe-1991}).

   The contribution of the mixing mechanism, that is of the interaction of different $\;^6T_{1u}\;$ CT terms of the octahedral [FeO$_6$]$^{9-}$ ($\;^6T_{2}\;$ CT terms of the tetrahedral [FeO$_4$]$^{5-}$) cluster can be written as follows\,\cite{Fe-1991})
\begin{widetext}
 \begin{eqnarray}
{\bf g}_a^{mix}\,=\sum_{{{j,k=^6T_{1u}}\choose{E_{0j}>E_{0k}}}}
 \frac{4\pi e^2 L N}{m_e}
\left(\lambda^{jk}\langle {\bf S} \rangle + \mu_B\,g_L^{jk}\,{\bf H}\right)
\left(\frac{f_j \,f_k}{\omega_{0j} \omega_{0k}}\right)^{1/2} \langle^6A_{1g}\| d \|j\rangle \langle ^6A_{1g}\| d \|k \rangle
 \frac{F_1(\omega,\,\omega_{0j})- F_1(\omega,\,\omega_{0k})}{E_{0j}-E_{0k}}\>,
\label{eq:52}
\end{eqnarray}
\end{widetext}
where $\langle^6A_{1g}\| d \|j\rangle$ is the dipole moment submatrix element.
The parameters of the type of effective
orbital $g$-factors
$\;g_L^{jk}\;$ and spin-orbit coupling constants $\; \lambda^{jk}\;$
\begin{eqnarray}
g_L^{jk}\,=\,\frac{\langle\kappa_j\>^6T_{1u}\|\sum_{n}{\bf l}_n\|\kappa_k\>^6T_{1u}\rangle}
{\langle 1 \|\,\hat l\, \| 1 \rangle}\>;\,\,\,
         g_L \equiv g_L^{jj} \equiv g_L^{j} \>;  \label{eq:53}
\end{eqnarray}
\begin{eqnarray}
\lambda^{jk}\,=\,\frac{\langle\kappa_j\>^6T_{1u}\|\,\hat Q^{11}\,\|\kappa_k\>^6T_{1u}
\rangle}{\langle 1 \|\,\hat l\,\| 1 \rangle \langle \frac{5}{2} \| \hat s \|\frac{5}{2} \rangle}\>; \,\,\,
       \lambda \equiv \lambda^{jj} \equiv \lambda^j\>  , \label{eq:54}
       \end{eqnarray}
are determined by the submatrix elements of the sum
$\; \sum_{n} {\bf l}_n \;$
of one-particle orbital moment operators acting on all atomic orbitals in the molecular orbitals, and by the submatrix element of the double irreducible spin-orbit
tensor operator $\; \hat Q^{11}\,$\,\cite{Griffith}. Numerical values of $\; g_L^{jk}\;$ and
$\; \lambda ^{jk}\;$  for the $\>CT\>$ states of the [FeO$_6$]$^{9-}$ and
[FeO$_4$]$^{5-}$
clusters are given in $Tables$ 1 and 2\,\cite{Fe-1991}.
In (\ref{eq:53}), (\ref{eq:54}), both the splitting $\;(j=k)\;$
and mixing $\; (j \not= k)\;$ are taken into account. $\>\kappa_j\;$ is the set of
 intermediate quantum numbers, necessary for distinguishing different
 $\; ^6T_{1u}\; $   terms. $\> f_j $  is the oscillator strength of the
$^6A_{1g} \rightarrow \kappa_j\>^6T_{1u}$ CT transition,
$ \; E_{0j}\; $   is its energy.

   Thus, $\; V_{SO}\;$  and $\; V_Z\; $ to the 1st order of the perturbation
   theory,  give rise to $isotropic$
$\;\hat A,\,\hat C\,$ tensors (\ref{eq:35}). The frequency
 dependences of the real
and imaginary parts of the splitting contribution to $\; {\bf g} \>$ for a
$\> CT\> $ transition
have respectively the "dissipative" and "dispersive" form.

The splitting contribution of the exchange-relativistic interaction $V_{SO}^{ex}$ (\ref{SoO})  for isolated ${}^6T_{1u}$ term to the gyration vector can be represented as follows\,\cite{Krich,Fe-1990,Fe-1991,Fe-1993}:
\begin{equation}
	{\bf g}=\frac{2\pi L e^2f_{AT}}{m\hbar\omega_0}
	\left(\stackrel{\leftrightarrow}{\pmb{\lambda}}\langle{\hat {\bf S}}\rangle + \sum_{n} \stackrel{\leftrightarrow}{\pmb{\lambda}}_{n}\langle{\hat {\bf S}}_{n}\rangle \right)\frac{\partial F(\omega ,\omega_0)}{\partial\omega_0}\, ,
	\label{gg0}	
\end{equation}
where first and second terms in brackets correspond to intra-center and inter-center, or spin-other-orbit exchange-relativistic contributions, respectively, $\stackrel{\leftrightarrow}{\pmb{\lambda}}$  and $\stackrel{\leftrightarrow}{\pmb{\lambda}}_{n}$ are the effective tensors of the respective interactions. In other words, these terms correspond to contributions with $m=n$ and $m\not=n$ in $V_{SO}^{ex}$ (\ref{SoO}). The summation over $n$ in (\ref{gg0}) extends to the nearest neighbors of the considered center,  $f_{AT}$ is the oscillator strength  of the ${}^6A_{1g}-{}^6T_{1u}$ transition.
In general, in accordance with (\ref{SoO}) the tensors $\stackrel{\leftrightarrow}{\pmb{\lambda}}$ and $\stackrel{\leftrightarrow}{\pmb{\lambda}}_{n}$ of the intra- and
 inter-center exchange-relativistic contributions in (\ref{gg0})  contain isotropic, antisymmetric, and symmetric anisotropic components.

 In addition to the "gyroelectric" contribution to the gyration vector that we have considered, we should note the existence of a small "gyromagnetic" contribution related with the magnetic susceptibility, which determines the frequency-independent contribution to the Faraday rotation\,\cite{Krinchik_UFN}
\begin{equation}
\Delta\Theta_F	= \frac{2\pi n_0}{c}\gamma\, m \, ,
\label{gg00}	
\end{equation}
where $\gamma$ is gyromagnetic ratio,  $m$ is magnetic moment. It is interesting that yttrium iron garnet in the wavelength range $\lambda >$\,5\,$\mu m$ is a gyromagnetic medium, since the gyromagnetic contribution to the Faraday rotation is predominant ($\Theta_F\approx$\,60\,deg/cm at T=300\,K), although in the wavelength range $\lambda <$\,4\,$\mu m$ it can be considered as an ordinary gyroelectric medium due to a sharp increase in the gyroelectric contribution in $\Theta_F$\,\cite{Krinchik_UFN}.

%\subsection{Circular magneto-optics in iron garnets}

\subsection{Fe$^{3+}$ diluted nonmagnetic compounds}

  The most suitable objects for the application and justification
of the cluster theory for ferrites are the Fe$^{3+}$ diluted nonmagnetic compounds such as YAlO$_3$ and Ca$_3$Ga$_2$Ge$_3$O$_{12}$ with the crystal structure close to  orthoferrite YFeO$_3$ and iron garnet Ca$_3$Fe$_2$Ge$_3$O$_{12}$, respectively. In such dilute systems, band models are inapplicable for describing Fe\,3d states, so that the cluster model has virtually no competitors in describing the optical and magneto-optical response of dilute systems in the O\,2p-Fe\,3d charge transfer range,  especially since it becomes possible to restrict ourselves to taking into account only intra-center p-d transfer.

The Faraday effect was measured in single-crystalline samples of diluted garnet Ca$_3$Ga$_{2-x}$Fe$_x$Ge$_3$O$_{12}$ ($x$\,=\,0.15)\,\cite{Fe-1990}, where the Fe$^{3+}$ ions occupy only the octahedral positions, and the [FeO$_6$]$^{9-}$ octahedrons are assumed to be essentially noninteracting. Making use of the splitting (\ref{eq:51}) and mixing (\ref{eq:52}) contributions to the gyration vector with the data for effective orbital $g$-factors and spin-orbital parameters from Table\,1 in Ref.\,\cite{Fe-1991} and assuming that energies of all "octahedral" CT transitions in
this garnet are blue-shifted by 1.4\,eV in comparison with corresponding
energies in "orthoferrite" complexes (see Table\,\ref{tableCT}), the authors calculated both the ferromagnetic and field contributions to the Faraday rotation \begin{eqnarray}
\Theta_F\,=\frac{\omega}{2\,n_0\,c}\,g\,=\, A_F\, m\, +\, C_F\, H\>,
\label{eq:89}
\end{eqnarray}
over the entire CT band. As a result, good agreement was obtained with the experimental values of the ferromagnetic and field contributions to $\Theta_F$, measured in the spectral range 1.4-3.1\,eV (see Figure\,2 in Ref.\,\cite{Fe-1990}).

Unfortunately, there are few examples in the literature of a systematic study of the concentration dependence of optical and magneto-optical effects in diluted systems.

\subsection{The yttrium iron garnet}

The absence of the magneto-optically active
rare-earth sublattice in yttrium iron garnet Y$_3$Fe$_5$O$_{12}$ permits the evaluation of the "undistorted" iron sublattices contribution. In addition, experimental studies of
$\; YIG \;$ magneto-optics are abundant \cite{Clogston,Kahn,Dillon,Scott}.
The authors\,\cite{Fe-1991} have undertaken a theoretical model computation of the FM and
field contributions (\ref{eq:35}) to the gyration vector of the YIG, taking into account the CT transitions
both in octahedral [FeO$_6$]$^{9-}$ and tetrahedral [FeO$_4$]$^{5-}$
clusters.

\begin{figure}[t]
	\centering
	%\sidecaption
		\includegraphics[width=8.5cm,angle=0]{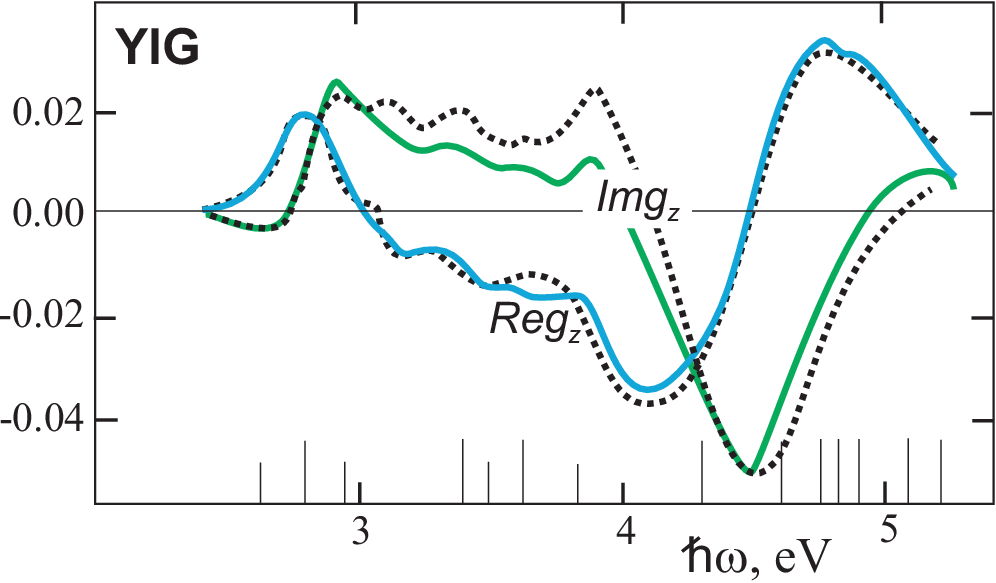}
		\caption{\footnotesize Spectral dependence of the real and imaginary parts of the $z$-component of the gyration vector in YIG: experimental data are shown by dotted curves, model fitting is shown by solid curves.}
	\label{fig5}
\end{figure}

Figure\,\ref{fig5} shows the results of  the theoretical simulation of the spectral dependence
of the real and imaginary parts of the gyration vector $\; z $ -component,
$\; Re\,g_z \;$ and $\; Im\,g_z \;$, in YIG (solid lines), with dipole allowed
and a number of dipole forbidden
CT transitions (marked by long and short line segments at the bottom of the
Figure\,\ref{fig5}, taken into account. The parameters of the main  CT
transitions used in the model
simulation are presented in Table\,\ref{tableCT}. Besides a satisfactory agreement with
the experimental data in a wide spectral range, 2.5 -- 5.5\,eV, the
computed  $\;Re\,g_z \;$ value on the long wavelength tail of the $\,CT \,$
transitions band $\; (\lambda = $ 0.63 $\, \mu m)\;$ yields the Faraday rotation in YIG
$\; \Theta_F \;  $ = 860 $ deg/cm\,$, practically
coinciding with the experimental value 830 $ deg/cm $\,\cite{Zvezdin,Hansen}.
The computed values of the partial Faraday rotation contributions due to
octahedral CT transitions (6500 $deg/cm $) and tetrahedral ones
(-\,5640 $deg/cm $) satisfactorily agree
with the experimental values 8670 and $-\,7840 \>deg/cm $, respectively
\cite{Zvezdin,Hansen}. As expected for a longitudinal ferrimagnet, we see the effect of significant mutual compensation for the contributions of the octa- and tetra-sublattices.

Both in the octahedral CT transitions contribution to $\,Re\>g_z\,$, and
in that of the tetrahedral transitions, the main role belongs to the mixing
mechanism, in agreement with the predominance of paramagnetic-shaped lines in
magneto-optical spectra of YIG noted in Ref.\,\cite{Scott2}.

The authors\,\cite{Fe-1991} have also computed  the field contribution (\ref{eq:35}) to
the YIG
gyration vector ${\bf g}$, with theoretical values of the orbital Land\'e
factors $\,g_L^{jk}\,$ (see Tables\,1 and 2 in Ref.\,\cite{Fe-1991}), taking into account the main
electric-dipole-allowed CT transitions, only. Rough as it is, the
approximation of allowed CT transitions gives nevertheless the
$\,\Theta_F/H\,$ values of $-\,10^{\circ}\,\cdot\,cm^{-1}\,\cdot\,T^{-1}\quad
(\lambda\,=\,0.7\>\mu m)\,$ and $\,-\,2.4^{\circ}\,\cdot\,cm^{-1}\,\cdot\,T^{-1}
\quad(\lambda\,=\,1.1\>\mu m)\,$ -- near to corresponding experimental data
($-\,12.4^{\circ}\,\cdot\,cm^{-1}\,\cdot\,T^{-1}\,$\,\cite{Kharchenko} and
 $-\, 2.5^{\circ}\,\cdot\,cm^{-1}\,\cdot\,T^{-1}\,$\,\cite{Berdennikova},
respectively). The lack of experimental data precluded a comparison at shorter wavelengths.

The electronic structure, magnetic, optical and magneto-optical properties of yttrium iron garnet were investigated recently\,\cite{Nakashima} by using "first principles" GGA+U
calculations with Hubbard energy correction for the treatment of the strong electron correlation.
The authors boldly make a too strong statement that "the calculated Kerr spectrum which included on-site
Coulomb interaction of Fe\,3d electrons described well the experimental results", which clearly does not follow from the data presented in Figure\,6 from their article, especially since the calculated dielectric function shows a dramatic discrepancy with experiment.

\subsection{Bi-substituted iron garnets}

Although pure yttrium iron garnet
has several advantages in terms of magneto-optical response, it has
not been widely applied in integrated devices due to its limited Faraday rotation.
However, decompensation of the contributions of the octa- and tetra-sublattices, in particular, due to the replacement of R-ions in R$_3$Fe$_5$O$_{12}$ garnets by Bi$^{3+}$ or Pb$^{3+}$ ions, makes it possible to increase the Faraday rotation of iron garnets by one or two orders of magnitude in the visible and near-infrared region (see, e.g., Ref.\,\cite{Wittekoek}).

Wittekoek et al.\,\cite{Wittekoek} proposed in a
purely qualitative manner that the origin of the large Faraday rotation in Bi,Pb-substituted iron garnets is the hybridization of Bi,Pb\,6p orbitals, which possess anomalously large
spin-orbit coupling ($\zeta_{6p}\approx$\,2\,eV), with the O\,2p and Fe\,3d orbitals. Later this idea was supported and developed within cluster molecular orbital theory\,\cite{Shinagawa,BIG-1991,BIG-2002}.
The enhancement of spin-orbit coupling in
Fe\,3d orbitals was assumed to be much smaller than that in O\,2p orbitals, because Fe sites are
located more distant than O sites from Bi sitess.

Taking account of the overlap of $\,2p\,(O^{2-})\,$ and $\,6p\,(Bi^{3+})\,$
electronic shells as well as the {\it virtual} transition of the oxygen $\,2p\,$-
electron to the bismuth empty $\,6p\,$- shell, the wave function of the outer
$\,2p\,$- electrons of the neighboring oxygen ion acquires thereby an admixture of Bi\,6p-states\,\cite{BIG-1991,BIG-2002}:
\begin{equation}
\varphi_{2p\,m}\,\longrightarrow\,\psi_{2p\,m}\>=\>\varphi_{2p\,m}\>-\>
\sum_{m^{'}}\,\langle 6p\>m'\,|\,2p\>m\rangle^{\ast}\,\varphi_{6p\,m'}\>\, ,
\label{eq:90}
\end{equation}
where $\,\varphi_{2p\,m}\,$ and $\,\varphi_{6p\,m}\,$ are atomic wave
functions.

The Bi\,6p-O\,2p hybridization
results in the modification of the
spin-orbit interaction on the oxygen ion :
\begin{equation}
V_{SO}\;= \;V_{SO}(2p)\; +\; \Delta V_{SO}^{iso}(2p)\; +
\; \Delta V_{SO}^{an}(2p)\> ,\label{eq:91}
\end{equation}
where
$V_{SO}(2p)=\zeta_{2p}\>({\bf l}\cdot{\bf s})$ is conventional spin-orbital interaction with $\zeta_{2p}\approx$\,0.02\,eV,
$\Delta V_{SO}^{iso}(2p)$ and  $\Delta V_{SO}^{an}(2p)$ are effective isotropic and anisotropic terms due to the Bi\,6p-O\,2p hybridization:
\begin{equation}
    \Delta V_{SO}^{iso}(2p)\; =\; \Delta \zeta_{2p}\>({\bf l}\cdot{\bf s})  \, ,
\end{equation}
where effective spin-orbital parameter is estimated in Ref.\,\cite{BIG-2002} to be $\Delta \zeta_{2p}\leq$\,0.1\,eV per one Bi$^{3+}$-ion, that is several times larger than conventional parameter $\zeta_{2p}$:
\begin{equation}
\Delta V_{SO}^{an}\,(2p)=\lambda_{ij}\, \hat l_i\, \hat s_j \>,
\label{eq:94}
\end{equation}
where the effective spin-orbit interaction tensor $\,\lambda_{ij}\,$ depends on the geometry of the Bi-O bond\,\cite{BIG-1991,BIG-2002}
\begin{equation}
\lambda_{ij} \propto \zeta_{6p}\left(R_i\,R_j\> - \>\frac{1}{3}\,\delta_{ij} \right) \, ,
\end{equation}
where ${\bf R}$ is a unit vector along the Bi-O bond direction.

Thus, the effect of the bismuth ions on the circular MOE in iron garnets is
essentially related to the oxygen O\,2p-states in [FeO$_6$]$^{9-}$ and
[FeO$_4$]$^{5-}$ clusters.
The Bi$^{3+}$ ions, leading to an increase in the effective spin-orbital coupling constant for oxygen ions, have a significant effect on the circular magneto-optics of iron garnets, through a change in the effective spin-orbital coupling parameters
$$\lambda = \lambda (3d)\> + \> \lambda  (2p)
$$
for the excited ${}^6T$-states with the p-d charge transfer.

The simple theory we are considering allows us to make a number of predictions. First, the effect of the Bi\,6p-O\,2p hybridization may be particularly significant for the
CT transitions, whose final state spin-orbit coupling constant
$\;\lambda\;$ contains the ligand contribution $\;\lambda(2p)\;$ only, e.g., the transitions $t_{2u}-e_g$ and $t_{1u}(\pi)-e_g$ in the [FeO$_6$]$^{9-}$ clusters (predicted energies 4.4 and 5.3\,eV, respectively). Since
$\zeta_{2p}\ll \zeta_{3d}\approx$\,0.1\,eV the contribution of such
transitions to the $\;FM\;$ part of the gyration vector (\ref{eq:35})
in unsubstituted garnets is practically vanishing. The Bi substitution
makes these transitions observable.
On the contrary, the CT transitions whose final state V$_{SO}$
constant $\lambda$ includes only the 3d-contribution, e.g.,
transition $t_{1u}(\sigma)-e_g$ in the [FeO$_6$]$^{9-}$ clusters (predicted energy 6.4\,eV) are not appreciably influenced by the Bi$^{3+}$-ions. Thus, the spectral dependence of the gyration vector in YIG and Bi-substituted compounds can differ greatly.
Second, the Bi\,6p-O\,2p hybridization induces the anisotropy of the $\hat A$ tensor in the FM contribution to the gyration vector (\ref{eq:35}), which differs for the octa- and tetra-positions of the Fe clusters.
Third, in our model, bismuth ions do not directly affect the value of the field contribution $\hat C\,{\bf H} $ (\ref{eq:35}) to the gyration vector.

At variance with the cluster model, the "first-principles" band  calculations indicate a slightly different, albeit contradictory, picture of Bi\,6p-O\,2p-Fe\,3d hybridization. Thus, analyzing the electronic structure of Bi$_3$Fe$_5$O$_{12 }$ (BIG) calculated by the fully relativistic first-principles method based on the full-potential linear-combination-of-atomic-orbitals (LCAO) approach within the local-spin-density-approximation (LSDA),
Oikawa et al.\,\cite{Oikawa} found that the enhancement of the spin-orbit coupling due to the hybridization of Bi\,6p is considerably larger in the Fe\,3d conduction bands than in the O\,2p and Fe\,3d valence bands. The origin of this
enhancement is that the Fe\,3d conduction bands energetically overlap with Bi\,6p bands. Their results indicate the significance of spin-orbit coupling in Fe\,3d conduction bands in relation to the large magneto-optical effect observed in BIG.

However, the results of recent GGA+U calculation by Li et al.\,\cite{Li} show that quite the contrary, Bi\,6p orbitals
in BIG hybridize significantly with Fe\,3d orbitals in the lower conduction bands, leading to large
V$_{SO}$-induced band splitting in the bands. Consequently, the  transitions between the upper valence bands and lower conduction bands are greatly enhanced when Y is replaced by Bi.
Such contradictions turn out to be typical for various "ab-initio" DFT based calculations.

\subsection{Exchange-relativistic interaction and unconventional magnetooptics of weak ferromagnetic orthoferrites}

Interestingly that circular magnetooptic effects in weak ferromagnets are anomalously large and are comparable with the effects in ferrite garnets despite two-three orders of magnitude smaller magnetization\,\cite{Kahn,Tabor,Zubov1,Chetkin,Zubov,Zubov-1992,Edelman}. In 1989 the anomaly has been assigned to a novel type of magnetooptical mechanisms related with exchange-relativistic interactions, in particular, with so-called spin-other-orbit coupling\,\cite{Krich}.

%\begin{displaymath}
%\mu \qquad
%\pmb{\lambda}
%\end{displaymath}

We have shown that an antisymmetric exchange-relativistic spin-other-orbit coupling
gives rise to an unconventional "antiferromagnetic" contribution to the circular magnetooptics for weak ferromagnets which can surpass conventional "ferromagnetic" term\,\cite{Krich,Fe-1990,Fe-1991,MO-2,MO-3,MO-4,Fe-1993} (see, also Ref.\,\cite{Zubov}).

The  gyration vector in weak ferromagnets is a sum of so-called ferromagnetic and antiferromagnetic terms with identical transformation properties, see Exp.\,(\ref{eq:35}).
It should be noted that within the two-sublattice model for orthoferrites we neglect weak antiferromagnetic A- and C-modes (see, e.g., Refs.\,\cite{JMMM-2016,CM-2019,JETP-2021,M-2021}).

For the first time the antiferromagnetic contribution to circular MOE was experimentally identified and evaluated in orthoferrite YFeO$_3$\,\cite{Krich}. An analysis of the field dependence of the Faraday rotation $\Theta_F({\bf H}_{ext})$  made it possible to determine all the contributions to the gyration vector ($\lambda$\,=\,0.6328\,$\mu$m):
$$
	A_{zz}m_z=(0.95\pm 0.55)\cdot 10^{-3};\, B_{zx}|l_x|=(3.15\pm 0.55) )\cdot 10^{-3};\,
	$$
	$$
A_{xx}m_x=(0.2\pm 0.7)\cdot 10^{-3};\, B_{xz}|l_z|=(-2.1\pm 1.0) )\cdot 10^{-3};\,
$$
\begin{equation}
C_{zz}\approx C_{xx}=(-1.1\pm 2.8) \cdot 10^{-6}\,kOe^{-1}\, ,
\end{equation}
where $|l_x|\approx |l_z|\approx 1$.
Interestingly, rather large measurement errors allow for certain to determine only the fact of a large if not a dominant antisymmetric antiferromagnetic contribution related with antisymmetric spin-other-orbit coupling.
Strictly speaking, the mutual orientation of the ferro- (${\bf m}$) and antiferromagnetic (${\bf l}$) vectors  depends on the sign of the Dzyaloshinskii vector\,\cite{JMMM-2016,CM-2019,JETP-2021}. Interestingly, a rather arbitrarily chosen relative orientation of these vectors in Ref.\,\cite{Krich}  with  positive sign of $m_z$ and $l_x$  exactly matches the theoretical predictions about the sign of the Dzyaloshinskii vector\,\cite{JMMM-2016,CM-2019,JETP-2021}.

\begin{figure}[t]
	\centering
	%\sidecaption
		\includegraphics[width=8.5cm,angle=0]{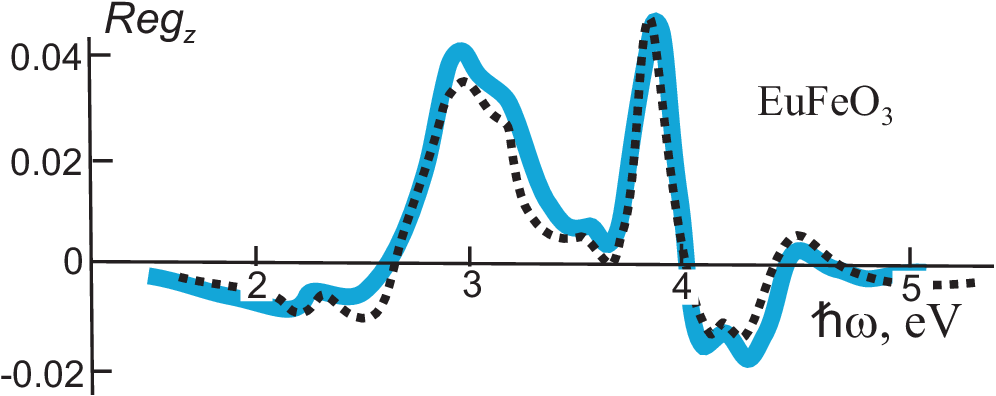}
		\caption{\footnotesize Spectral dependence of the real part of the $z$-component of the gyration vector ib EuFeO$_3$: experimental data are shown by dotted curve, model fitting is shown by solid curve.}
	\label{fig6}
\end{figure}

Existence of spontaneous spin-reorientational phase transitions $\Gamma_4(F_zG_x)\rightarrow\Gamma_2(F_xG_z)$ in several rare-earth orthoferrites does provide large opportunities to study anisotropy of circular magnetooptics\,\cite{Kahn,Tabor,Chetkin,MO-2,MO-3,MO-4,Fe-1993}. Gan'shina {\it et al.}\,\cite{MO-3} measured the equatorial Kerr effect in EuFeO$_3$, TmFeO$_3$, and HoFeO$_3$ and have found the the gyration vector anisotropy in a wide spectral range 1.5-4.5\,eV. The magnetooptical spectra, both real and imaginary parts of the gyration vector, were nicely fitted within a microscopic model theory based on the dominating contribution of the O2$p$--Fe3$d$ charge transfer transitions and spin-other-orbit coupling  in [FeO$_6$]$^{9-}$ octahedra. An example of modeling the spectrum of the real part of the gyration vector in orthoferrite EuFeO$_3$ is shown in Figure\,\ref{fig6}. Let us again pay attention to the comparable values of circular MOEs in orthoferrites and ferrite garnets at more than an order of magnitude lower magnetic moment in weak ferromagnets of the YFeO$_3$ type and longitudinal ferrimagnets of the YIG type. The authors\,\cite{MO-3} have demonstrated a leading contribution of the antisymmetric spin-other-orbit coupling and  estimated effective orbital magnetic fields in excited $^6T_{1u}$ states of the [FeO$_6$]$^{9-}$ octahedra, $H_L\sim 100\,T$. These anomalously large fields can be naturally explained to be a result of strong exchange interactions of the charge transfer $^6T_{1u}$ states with nearby octahedra that are determined by a direct $p$\,-\,$d$ exchange.

Whereas the existence of the antiferromagnetic contribution to the gyration vector is typical of a large number of multisublattice magnetic
materials, the antisymmetry of the tensor $\stackrel{\leftrightarrow}{B}$  is a specific feature
of weak ferromagnets alone. In the case of rhombohedral weak ferromagnets such as FeBO$_3$, FeF$_3$, or $\alpha$-Fe$_2$O$_3$,
the tensor $\stackrel{\leftrightarrow}{B}$, governing the antiferromagnetic contribution
to the Faraday effect is entirely due to the antisymmetric
contribution, in view of the requirements imposed by the
crystal symmetry. In crystals of this kind the appearance of
the antiferromagnetic contribution to the gyration vector is
entirely due to allowance for the antisymmetric spin-other-orbit coupling.

However, the data on the anisotropy of the Faraday effect in TmFeO$_3$
\cite{Chetkin} and
the values of the Faraday effect in SmFeO$_3 \quad ({\bf m} \parallel a $-axis)
and a number of other orthoferrites with $\; {\bf m} \parallel c $ -axis
\cite{Tabor}
bear evidence of the existence of an appreciable symmetric $\; AFM \quad
\hat B^{s} {\bf l} \;$ contribution to the gyration vector of orthoferrites.
Indeed, the Faraday effect in the $\; \Gamma_4 \,$ phase $\;({\bf m} \parallel
 c) \;$ and in the $\; \Gamma_2 \,$ phase $\; ({\bf m} \parallel a) \;$ is
 determined, respectively, by the $\> z$-  and $\>x\,$- component of $\;{\bf g}\>$:
\begin{equation}
g_z \> = \> A\,m_z \> + \> B_{zx}\,l_x\,;\quad g_x \>= \> A\, m_x \>+\>B_{xz}\,l_z
\label{eq:37}
\end{equation}
(under the justified assumption that $\> \hat A \>$ be isotropic). Since
$ \; {\bf m} \, \perp \, {\bf l} \;$ and $\; m_x \, \approx \, m_z \, = \, m \,$,
letting $\; l_x \, = \, 1\;$ with the view of the definitude, we obtain :
\begin{equation}
g_z \,=\, A\,m\,+\,B_{zx}^{a} \,+ \, B_{zx}^{s}; \quad
g_x \,=\, A\,m\,+\,B_{zx}^{a} \,- \, B_{zx}^{s} \; ,\label{eq:38}
\end{equation}
so that the experimentally found ratio \cite{Chetkin,Tabor} $\; Re\,g_z \,/ Re \, g_x
\, \approx \,$ 2.5$\, - \,$3  (at $\; \lambda \, \approx \, $ 1$\,-\,$2$\,
\mu m\,$) indicates unambiguously the existence of an appreciable
symmetric $\,AFM\,$ term $\,B_{zx}^{s}$:
\[
\frac{B_{zx}^{s}}{A\,m\,+\,B_{zx}^{a}}\,\sim \,0.5\>.
\]

\subsection{The temperature dependence of the circular magneto-optics of ferrites}

The analysis of the temperature dependences of $\,MOE\,$ can yield an
important information about the role of various mechanisms of the circular
$\,MOE\,$.
Experimental studies of the Faraday and Kerr effects in weak ferromagnets
$\,\alpha$-Fe$_2$O$_3$ \cite{Zubov},$\;FeBO_3\,$ \cite{Edelman,Zubov-1992},YFeO$_3$ \cite{Zubov1} have shown that their circular $\,MOE\,$ and
the magnetic moment, both total and that of each sublattice,  have
{\it different} temperature dependences. In Refs.
\cite{Zubov,Zubov1,Edelman}, an attempt was made to
connect this phenomenon with the so-called $\>pair\>$ transitions.

   However, we show here that all peculiarities of the temperature dependence of
the Faraday and Kerr effects for weak ferromagnets can be naturally and consistently explained
by taking into account the $\>AFM \>\hat B {\bf l}\;$ contribution to the
gyration vector due
to the exchange-relativistic interactions. Whereas the $\;FM \> \hat A
{\bf m}\;$ contribution to $\;{\bf g}\;$ (\ref{eq:35}), the
$\;AFM \> \hat B^{sym}{\bf l}\;$ contribution
due to $\;LSCF\;$ in $\;^6T_{1u}\quad CT\;$ states (\ref{eq:35}),
and the contributions due to intra-center $\;V_{so}^{ex}\;$
have the
temperature dependence determined by the ordinary thermodynamic average
of the spin $\langle S(m) \rangle$, the $\,AFM\,$ contribution owing to
the "spin-other orbit" interaction is related to the average value of a complicated spin operator
$\,\tilde S(mn)\,$ (\ref{Smn}).
\begin{figure}[t]
	\centering
	%\sidecaption
		\includegraphics[width=8.5cm,angle=0]{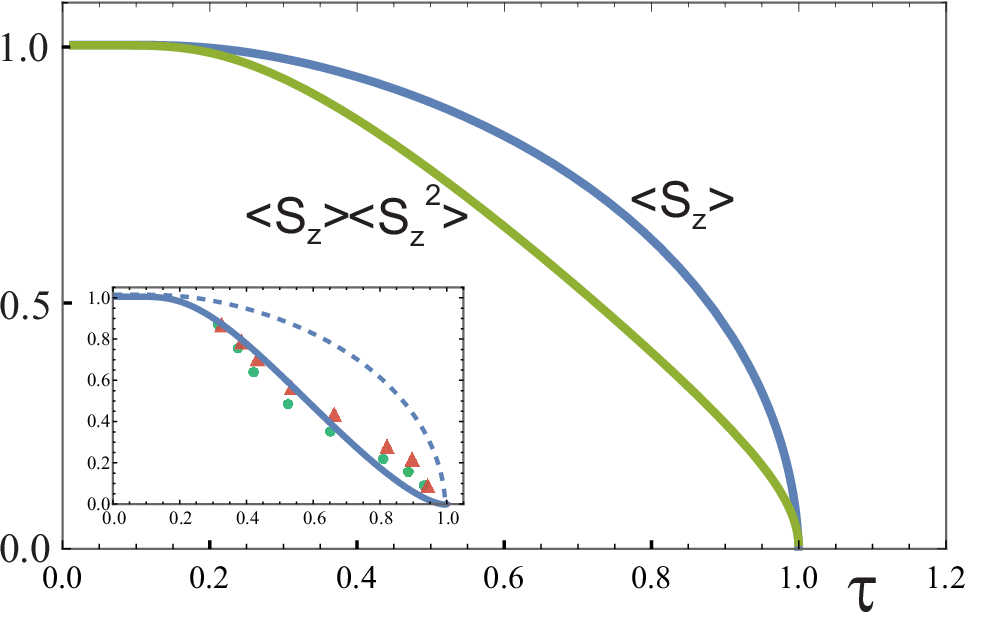}
		\caption{\footnotesize Temperature dependence of the normalized thermodynamic quantities determining the temperature dependence of the  circular MOE. The inset shows an example of fitting the experimental data on the temperature dependences of the equatorial Kerr effect in hematite $\alpha$-Fe$_2$O$_3$ (see Figure\,6 in Ref.\,\cite{Zubov1}) using the two-parameter formula (\ref{eq:888}), dotted curve is the $\langle S_z\rangle$ dependence.}
	\label{fig7}
\end{figure}
In the molecular field approximation the thermodynamic average of the nonlinear operator  ${\bf S}_{mn}$ in (\ref{Smn}) can be written as follows\,\cite{Callen}
\begin{widetext}
\begin{eqnarray}
\langle\hat S_q(mn)\rangle\,=\, \langle\hat S_z(n)\rangle C^1_q({\bf S}(n)) \,+\,
\gamma \langle\hat V^2_{0}(S(m))\rangle_T \langle\hat S_z(n)\rangle  \sum_{q_1,q_2}  \left[
\begin{array}{ccc}
2 & 1 & 1 \\
q_1 & q_2 & q
\end{array}\right] C^2_{q_1}({\bf S}(m)) C^1_{q_2}({\bf S}(n)),
\label{<Smn>}
\end{eqnarray}
\end{widetext}
where $C^2_{q_1}({\bf S}(m)), C^1_{q_2}({\bf S}(n))$ are  spherical tensorial harmonics ($C^k_q=\sqrt{\frac{4\pi}{2k+1}}Y_{kq}$) as the functions of classical spin direction;
$$
\langle S_z \rangle \;=\; S\,B_S (x)\, ,
$$
where $\;B_S (x)\;$ is the Brillouin function
$$
B_S (x)=\frac{2S+1}{2S}coth\frac{2S+1}{2S}x-
\frac{1}{2S}coth\frac{1}{2S}x\,;\,\,\, x\, =\, \frac{3S}{S+1}\,\frac{\sigma}{\tau}\;
$$
($\sigma\;=S_z/S$ and $\; \tau \;$
being the reduced magnetic moment and temperature, respectively);
\begin{eqnarray}
\langle\hat V^2_{0}(S)\rangle_T=2\,\left[\frac{(2S\,-\,2)!}{(2S\,+\,3)!}\right]^{1/2}
\left( 3 \langle\hat S_z^2\rangle \>-\> S(S \,+\,1) \right) \, ,
\end{eqnarray}
where
$$
\left( 3 \langle\hat S_z^2\rangle \>-\> S(S \,+\,1) \right)=
$$
\begin{eqnarray}
\left( 2S\,(S\>+\>1) -  3S\,\coth \frac{x}{2S} \cdot B_S (x)\right) \, ,
\end{eqnarray}
Thus, the temperature dependence of the gyration vector in the molecular field
approximation is determined by the following two-parameter formula :
\begin{equation}
     g(T) \>= \> a\,\langle S_z\rangle+\> a^{\prime}\,\langle S_z\rangle \langle S_z^2\rangle \approx A\,m+A^{\prime}\,m^3,
     \label{eq:888} \, ,
\end{equation}
with the frequency dependent coefficients $a,\> b$. Temperature dependence of the thermodynamic factors $\langle S_z\rangle$ and $\langle S_z\rangle \langle S_z^2\rangle$ are presented in Figure\,\ref{fig7}, where the inset shows an example of fitting experimental data on the temperature dependences of the equatorial Kerr effect in hematite $\alpha$-Fe$_2$O$_3$ (see Figure\,6 in Ref.\,\cite{Zubov1}) using the two-parameter formula (\ref{eq:888}).

In other words, the MOE in weak ferromagnets will be characterized by a clear nonlinear dependence on the magnetic moment of sublattices, the presence of which is a direct indication of the contribution of exchange-relativistic interactions of the spin-other-orbit type. As expected, the nonlinear contribution, both in magnitude and in sign, will depend substantially on the frequency\,\cite{Zubov,Zubov1,Zubov-1992,Edelman}.

It is worth noting that Exp.\,(\ref{<Smn>}) provides a dependence of the exchange-relativistic contribution to the gyration vector on the mutual orientation of neighboring spins.

%$Figure$ @@ shows results of the processing of experimental data \cite{Zubov,Zubov1} by means of the relation (\ref{eq:88}). The fit is quite good in all cases, what undoubtedly is an evidence for our conclusion of the importance of the exchange-relativistic interactions in $\,CT\,$ states for the magneto-optics of orthoferrites and hematite.

\subsection{The high-energy optics and magneto-optics of ferrites}

The availability of modern high-intensity synchrotron radiation has
facilitated the refinement of conventional spectroscopy. This is especially true in the field of
$\,MOE\,$, where the synchrotron radiation is a convenient tool of obtaining
the spectra at high energies.

Ku\v{c}era $et\>al.$\,\cite{Kucera90} have obtained the reflectivity
spectra of a number of iron and non-iron garnets and yttrium
orthoferrite in the vacuum ultraviolet 5 to 30\,eV range using synchrotron
radiation as the light source. Contrary to the visible and near UV regions, all the spectra obtained are strikingly similar in this spectral range.
Two broad bands sited at about 10 and 17\,eV have been found in both garnet and orthoferrite reflectivity and optic absorption spectra.
The 10\,eV band was assigned to the CT transition from the
oxygen 2p valence band  to the yttrium 4d or
5s conduction states. The band centered near 17\,eV was attributed to the "orbital-promotion"\, inter-configurational  Fe\,3d$\rightarrow$\,Fe\,4p transition.
Despite the large peak values, the contribution of these transitions to the MOE of ferrites in the visible region, being structureless, is significantly inferior to the contribution of O\,2p-Fe\,3d CT transitions.

\subsection{Rare-earth ions in ferrites}

The simplest expression for the contribution of the dipole-allowed 4f–5d transition to the rare-earth ion polarizability tensor can be obtained by neglecting the splitting of the 4f$^{n-1}$5d- configuration\,\cite{Pleshchev_88}
\begin{equation}
\alpha _q^k \,= \, (-1)^{1+k}3\sqrt{2k+1} \frac{1}{\hbar}
\left\{ \begin{array}{ccc}
3 & 3 & k \\
1 & 1 & 2
\end{array} \right\}
e^2r_{fd}^2 F_k(\omega, \, \omega_{fd})
\langle \hat U^k_q(J) \rangle
\label{eq:44}
\end{equation}
where \{:::\} is the 6j-symbol\,\cite{Sobelman}, $r_{fd}=\langle 4f|r|5d\rangle$ is the radial integral, $\langle \hat U^k_q(J) \rangle$ is the thermodynamical average of the  irreducible tensor $ \hat U^k_q(J)$ with submatrix element $U^{(k)}_{SLJ;SL^{\prime}J^{\prime}}$\,\cite{Sobelman}.

The components of the tensor $\alpha^1_q$, which determines the contribution of the rare-earth ion to the circular magneto-optics, can be written as follows
\begin{equation}
\bm{\alpha}  \,= \, -\frac{1}{7\sqrt{2}}
e^2r_{fd}^2 F_1(\omega, \, \omega_{fd})
\frac{2-g_J}{g_J\mu_B} {\bf m}_R \, ,
\label{eq:45}
\end{equation}
where ${\bf m}_R$ is the magnetic moment of the R-ion, $g_J$ is the Lande-factor.
The symmetric anisotropic part of the polarizability tensor determines the effects of linear birefringence and dichroism. In Cartesian form, we get\,\cite{Pleshchev_88}
\begin{equation}
 \alpha_{ij}  \,= \, \frac{\sqrt{3}}{14}
e^2r_{fd}^2 F_2(\omega, \, \omega_{fd})
\alpha \langle 3\widetilde{\hat J_i\hat J_j}-J(J+1) \rangle \, ,
\label{eq:45}
\end{equation}
where $\widetilde{J_iJ_j}=\frac{1}{2}(\hat J_i\hat J_j+\hat J_j\hat J_i)$, $\alpha$ is the Stevens parameter\,\cite{Taylor}.

A detailed analysis of the role of the effects of a strong crystal field for the 5d electron was carried out in Refs.\,\cite{Pleshchev_88,Pleshchev_90}.

\section{Conclusions}

The paper presents the theory of the optical and magneto-optical properties of strongly correlated iron oxides, primarily ferrite garnets and orthoferrites, based on the cluster model with the leading contribution of the charge transfer transitions.

At variance with the "first-principles" DFT based band models the cluster model is physically clear, it allows one to describe both impurity and dilute and concentrated systems, provides a self-consistent description of the optical, magnetic, and magneto-optical characteristics of Fe centers with a detailed account of local symmetry, low-symmetry crystal field effects, spin-orbit and Zeeman interactions, and also relatively new exchange-relativistic interaction, which plays a fundamental role for the circular magneto-optics of weak ferromagnets.

The cluster approach provides a regular procedure for classifying and estimating the probability of allowed and forbidden electric-dipole CT transitions and their contribution to optical and magneto-optical anisotropy.

The cluster model makes it possible to describe all the specific features of the influence of Bi ions on the circular magneto-optics of  ferrites by the Bi\,6p-O\,2p  hybridization and partial Bi-O "transfer" of the large Bi\,6p spin-orbit interaction. The cluster model predicts the "selective" nature of the influence of Bi only on certain CT transitions, the appearance of an anisotropy of the ferromagnetic contribution, and the absence of any influence on the field contribution to the gyration vector.

The contribution of the exchange-relativistic interaction for the excited ${}^6T_{1u}$ terms in [FeO$_6$]$^{9-}$ clusters leads not only to the appearance of an "antiferromagnetic" contribution to the gyration vector of weak ferromagnets such as orthoferrite RFeO$_3$ and hematite $\alpha$-Fe$_2$O$_3$ but also to the deviation of the temperature dependence of circular MOE from the simple proportionality to the magnetization $\bf m$. The appearance of a nonlinear $\bf m$-dependence is an indication of the contribution of the unusual  "spin-other-orbit" interaction in excited ${}^6T_{1u}$ states.

Undoubtedly, the considered version of the cluster theory requires more detailed development both in terms of improving the used MO-LCAO scheme and in terms of the possible application of the "hybrid"  LDA\,+\,MLFT scheme\,\cite{Haverkort}. In any case, development
 the cluster model of magneto-optical effects in ferrites needs data from systematic experimental studies of the concentration, spectral, and temperature dependences of various optical and magneto-optical effects for Fe centers in oxides.

%\funding{Supported by Russian Science Foundation, grant number 22-22-00682 (Secs.\,1-5, 7) and the Ministry of Education and Science project No FEUZ-2020-0054   (Sec.\,6).}

\begin{acknowledgments}
This study was supported by the Ministry of Science and Higher Education of the Russian Federation, project FEUZ-2023-0017
\end{acknowledgments}

%\ack This study was supported by the Ministry of Science and Higher Education of the Russian Federation, project FEUZ-2023-0017

\end{document}